\documentclass[manuscript]{aastex}

\usepackage{color}
\usepackage{lineno}
\usepackage{rotating}
\usepackage{lscape}
\usepackage{comment}
\usepackage{longtable}
\usepackage{lineno}
\usepackage{ulem}

\shorttitle{Origin of the UV to X-ray emission of radio galaxy NGC 1275 explored by analyzing its variability}
\shortauthors{F. Imazato et al.}

\begin{document}

\title{Origin of the UV to X-ray emission of radio galaxy NGC 1275 explored by analyzing its variability}

\author{
Fumiya Imazato\altaffilmark{1}, 
Yasushi Fukazawa\altaffilmark{1,2,3}, 
Mahito Sasada\altaffilmark{2},
Takanori Sakamoto\altaffilmark{4}
}

\altaffiltext{1}{Department of Physical Science, Hiroshima University, 1-3-1 Kagamiyama, Higashi-Hiroshima, Hiroshima 739-8526, Japan}
\altaffiltext{2}{Hiroshima Astrophysical Science Center, Hiroshima University, 1-3-1 Kagamiyama, Higashi-Hiroshima, Hiroshima 739-8526, Japan}
\altaffiltext{3}{Core Research for Energetic Universe (Core-U), Hiroshima University, 1-3-1 Kagamiyama, Higashi-Hiroshima, Hiroshima 739-8526, Japan}
\altaffiltext{4}{Department of Physics and Mathematics, College of Science and Engineering, Aoyama Gakuin University, 5-10-1 Fuchinobe, Chuo-ku, Sagamihara-shi, Kanagawa 252-5258, Japan}

\begin{abstract}
We analyze the ultraviolet (UV) and X-ray data of NGC 1275 obtained with {\it Swift}/UVOT, XRT, BAT and {\it Fermi} Large Area Telescope over about 10 years to investigate the origin of the nuclear emission from NGC 1275.
We confirm that the UV and soft/hard X-ray fluxes gradually increased along with the GeV gamma rays. 
At times, short-term variations in the UV or soft X-ray spectral regions showed rapid variations correlated with the GeV gamma-rays. 
However there was no significant correlation between the UV and soft X-rays. 
The UV spectrum had a narrow spectral shape that could be represented by single-temperature blackbody radiation.
These results could possibly indicate that the long-term variability of UV and X-ray emissions is caused by the jet, while the emissions from the accretion disk contribute to the UV and X-ray bands to some extent.
\end{abstract}

% \vspace*{1cm}

\section{Introduction}
A blazar is a class of AGN in which the jet points toward our line of sight. Conversely, a radio galaxy is a class of AGN in which the jet axis is misaligned with our line of sight.
If the jet of a radio galaxy has a spine-sheath structure \citep{Ghisellini2005A&A}, we can observe the outer part of the AGN jet.
Therefore, both blazars and radio galaxies are important for studying AGN jets. However, it is difficult to analyze the AGN emissions (jet and disk/corona) in optical and ultraviolet (UV) bands for radio galaxies because of host galaxy contamination, and it is also difficult to disentangle the emissions from the jet and disk/corona in optical/UV and X-ray bands.
Consequently, the origin of optical/UV and X-ray emissions is generally unclear for radio galaxies.
Therefore, it is necessary to study such emissions to study the AGN jet structure and the connection between the AGN jet and accretion disk.
NGC 1275 (3C 84; $z$ = 0.0176: \citet{Young1995}) is a cD galaxy at the center of the Perseus cluster.
It is classified as a radio galaxy or a radio-loud Seyfert galaxy.
The AGN jet activity of NGC 1275 restarted in 2005, and the radio flux has gradually increased since \citep{Nagai2016}.
Thus, it provides us with an observational opportunity to study the growth of the AGN jet.

Since 2008, NGC 1275 has been detected in the GeV ($>$0.1 GeV) gamma-ray band by the Fermi Large Area Telescope ({\it Fermi}/LAT), and it is the brightest radio galaxy in this energy range \citep{Abdo2009}.
The timescale of the GeV gamma-ray variation is a few months \citep{Kataoka2010}, and several flares have been detected \citep{Donato2010,Brown2011,Tanada2018}.
These results mean that the main origin of the GeV gamma-rays is not from the Perseus cluster itself.
In addition, NGC 1275 was also detected in the TeV ($>$ 0.1 TeV) gamma-ray band by the Major Atmospheric Gamma Imaging Cherenkov (MAGIC) telescopes \citep{Aleksi2012,Aleksi2014a} and the Very Energetic Radiation Imaging Telescope Array System (VERITAS) \citep{Benbow2015}.
The MAGIC team reported no, or, at most, weak (3.6$\sigma$), evidence of TeV gamma-ray variability on month-long timescales in two observational campaigns between 2009 October and 2010 February and between 2010 August and 2011 February.
Recently, the brightest TeV gamma-ray outburst was observed between December 31, 2016 and January 1, 2017, with a flux 50 times higher than that previously measured flux \citep{Magic2018}.
The VERITAS team also measured a very high TeV gamma-ray flux on January 2, 2017 \citep{VeritasATel2017} after the MAGIC report \citep{MagicATel2017}.

Optical continuum flux variations have also been detected \citep{Yuan2015,Pronik1999,Aleksi2014a,Magic2018}, and \citet{Pronik1999} found variations in the UV continuum flux over 37 nights from December 22, 1989, to December 29, 1994.
In addition, \citet{Aleksi2014a} have reported a positive flux correlation between the optical (R-band) and GeV gamma-ray band at the level of 4$-$5 $\sigma$. 
This strongly suggests that optical (R-band) emission is a counterpart to gamma-ray emission.
\citet{Martin1976} measured optical polarization on January 4$-$7, 1976, with the 2.3-m telescope at Steward Observatory and discovered variations on a timescale of one day.
However, \citet{Yamazaki2013} measured the optical continuum (6018$-$6205\AA) and polarization in 2010$-$2011 with the 1.5-m Kanata telescope of the Higashi-Hiroshima Observatory and reported the variability of the optical continuum flux to be less than 13\% and that the degree of polarization was very small, around 0.4\% with little variability; note that host galaxy contribution reduced the polarization degree in this case.

Long-term variation has been found in the soft X-ray band, together with radio and GeV gamma-ray variations \citep{Fabian2015,Fukazawa2018}.
These probably indicate that we observe a contribution from jet emission at X-ray energies.
However, a fluorescent Fe K line at 6.4 keV has also been detected by {\it XMM-Newton} and {\it Hitomi}/SXS \citep{Churazov2003,Yamazaki2013,Hitomi2018PASJ}.
Such a line is emitted when circumnuclear material is illuminated by non-beamed X-ray emission from a galactic nucleus, which requires disk/corona X-ray emission.

A one-zone synchrotron self-Compton (SSC) emission model has often been used in fitting the spectral energy distribution (SED) of the nucleus of NGC 1275 (e.g., \citealt{Abdo2009,Aleksi2014a,Suzuki2012}). As described above, however, the origin of the optical/UV and X-ray bands of NGC 1275 is still not well understood.
Thus, in this paper, we analyze the {\it Swift}/UVOT data for NGC 1275 using point-spread function (PSF) photometry covering the period from July 13, 2007, to April 4, 2017, to study the optical/UV variability over the longest period examined to date.
We also extend the analysis period for the {\it Swift}/XRT soft X-ray data from NGC 1275, originally presented by \citet{Fukazawa2018}, up to March 31, 2017. Furthermore, we analyze the {\it Swift}/BAT hard X-ray data for NGC 1275 from 2006 through 2017.
In addition, we use {\it Fermi}/LAT GeV gamma-ray data of NGC 1275 from Goddard Space Flight Center \footnote{$\rm https://fermi.gsfc.nasa.gov/ssc/data/access/lat/msl\_lc/$} to compare the gamma-ray emission from the jet with the optical/UV and X-ray band emissions. Throughout this paper, the errors we adopt for the {\it Swift}/XRT and BAT results correspond to the 90\% confidence level, and those for the {\it Swift}/UVOT and {\it Fermi}/LAT results correspond to one-sigma errors.
We describe the analyses of the {\it Swift}/UVOT, XRT, and BAT data in section 2. We describe our results in section 3 and discuss them in section 4. 
We summarize our conclusions in section 5.

\clearpage
%
%__________________________________________________________________
\section{Data reduction and analysis}

\subsection{UVOT}

{\it Swift}/UVOT is an optical and UV telescope. 
It has six filters (UVV, UBB, UUU, UVW1, UVM2, and UVW2, covering the wavelength range 170$-$650 nm), and the pixel size of all CCD images is 0.502 arcsec/pix. 
Herein, we used only data with datamode $=$ IMAGE, to which we applied  aspect and mod8 corrections.
We obtained the UVOT images from the UK Swift Science Data Centre.\footnote{$\rm http://www.swift.ac.uk/swift\_live/index.php$}
Since the AGN emission is embedded in the galaxy emission from NGC 1275, we derived the AGN flux using the PSF photometry technique. 
We fitted a one-dimensional profile to the CCD counts to separate the AGN emission from the galaxy emission.

\begin{figure}
    \centering
    \includegraphics[width=8.7cm]{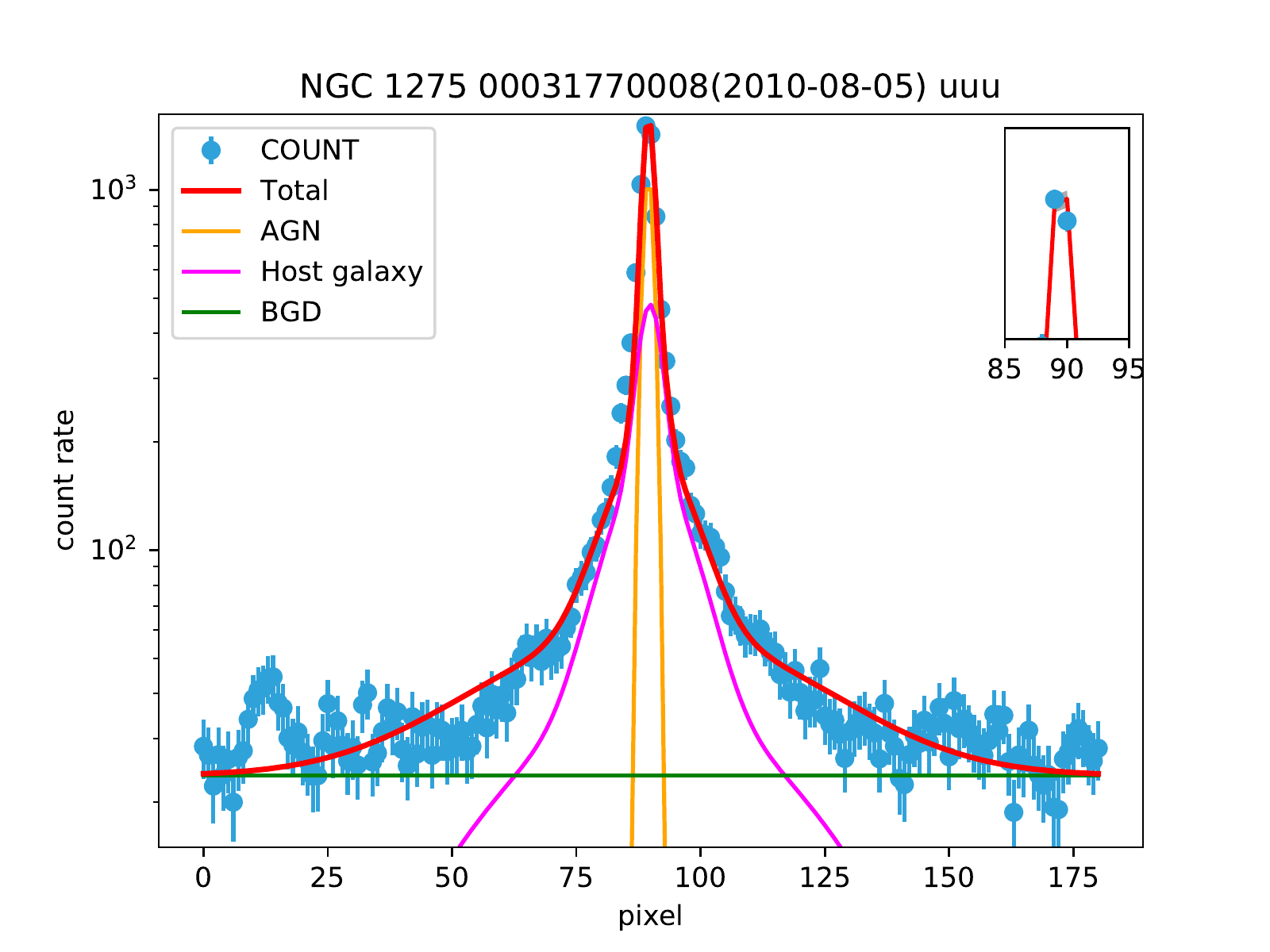}
    \caption{One-dimensional UVOT (UUU band) profile of NGC 1275, fitted by using the AGN and the host-galaxy model. The yellow line is the AGN emission model, the pink line is the host-galaxy model, the green line is the background (BGD) emission model (constant), and the red line is the AGN + host-galaxy + BGD emission model. One pixel is 0.502 arcsec.}
    \label{fig:NGC1275_fitting}
\end{figure}

To start, we obtained a summed image using {\tt UVOTSUM} for each color in every observation. 
We extracted a region with 11 $\times$ 181 pixels (5.522 arcsec $\times$ 90.862 arcsec) centered on NGC 1275 for each image. 
We call this the "{\it sum region}."
We then projected the photon counts from the {\it sum region} in one direction as shown in Fig. \ref{fig:NGC1275_fitting}.
To make a count-profile model of the parent galaxy, we extracted the count profile of NGC 1272, an elliptical galaxy close to NGC 1275, in the same way as described above.
We fitted it with three Gaussians plus a constant in all period images, and we used the median values of all the Gaussian parameters as the parameters for the host-galaxy model.
For the AGN count-profile model, we analyzed the star Pul $-$3 270315 ($\alpha$ = 03:19:41.74244, $\delta$ = +41:30:36.688) in the image, and we fitted it with one Gaussian plus a constant.
The parameters obtained are listed in Table \ref{tab:para_agn_and_hostgalaxy}. 
Fixing these parameters, we fitted one-dimensional count profiles for NGC 1275 with a linear combination of these AGN and galaxy models, together with a constant as a background model.
To start, we fitted the count profile of the {\it sum region} using only a host-galaxy model, excluding the AGN emission region (85 $<=$ pixel $<=$ 95 in Fig. \ref{fig:NGC1275_fitting}) and star emission region (0 $<=$ pixel $<$ 30 in Fig. \ref{fig:NGC1275_fitting}) to determine the host-galaxy model.
Then, we fitted the entire region of the count profile with an AGN model and the fixed host-galaxy model.

\begin{figure}
    \centering
    \includegraphics[width=8.7cm]{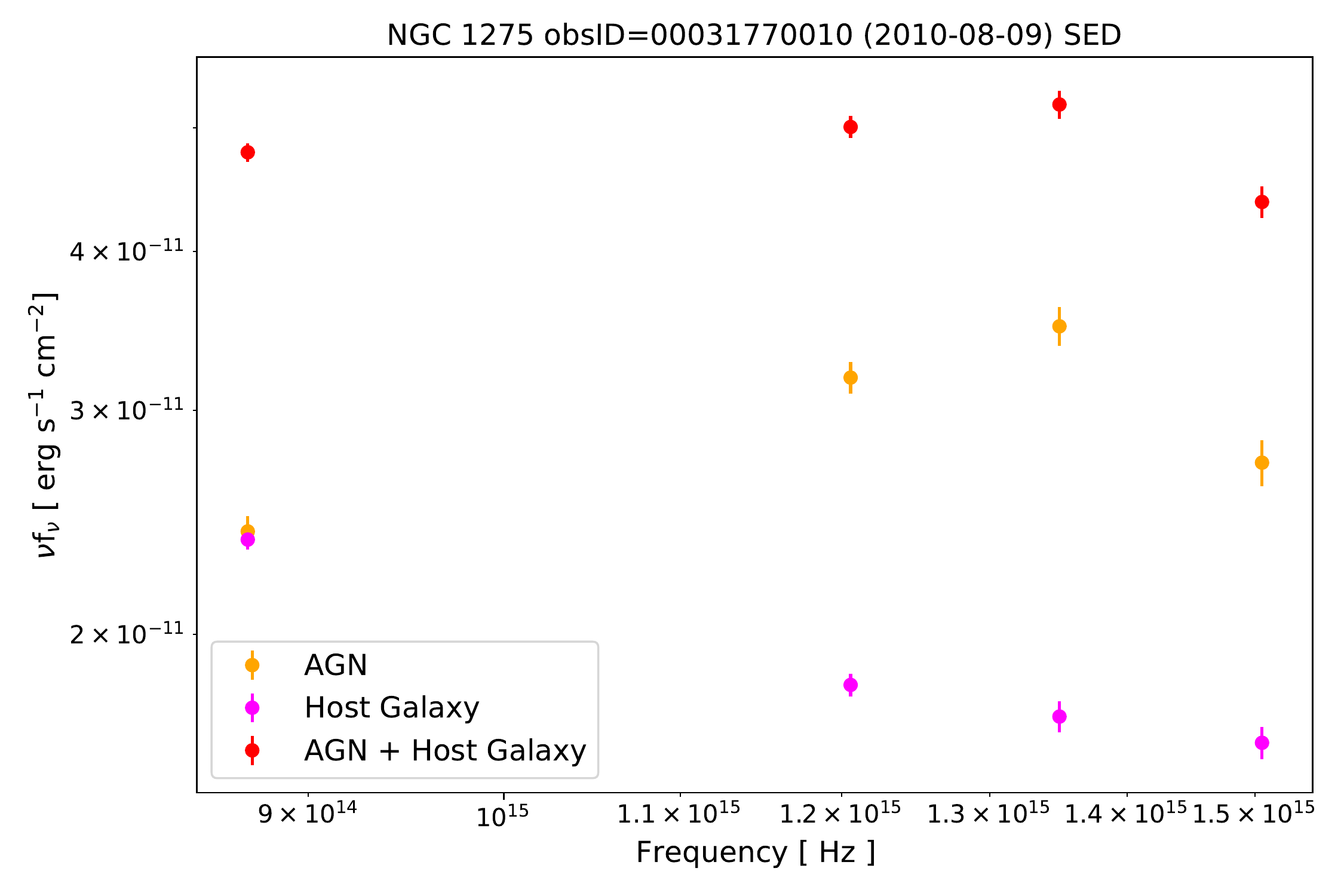}
    \caption{UV SED of NGC 1275, where the AGN emission and the host galaxy emission are separated. The observation ID (obsID) is 00031770010 (MJD = 55417.7377). The yellow, pink, and red circles are the AGN, host-galaxy, and AGN plus host-galaxy emission, respectively.}
    \label{fig:separation_agn}
\end{figure}

\begin{table}[hbtp]
    \caption{Best-fit parameters for the AGN and host-galaxy emission models from the UVOT one-dimensional count profile. Each of the three "components" in the host-galaxy emission model represents one of the three Gaussians to fit the host galaxy.}
    \label{tab:para_agn_and_hostgalaxy}
    \centering  
    \begin{tabular}[t]{lcc}
        \hline \hline
        Model & Gaussian sigma [ pixel ] & Gaussian normalization ratio \\ \hline
        AGN model & 1.143 & \\ \hline
        Host galaxy model & & \\
        Component 1 & 2.306  & 1 \\
        Component 2 & 28.796  & 1.389\\
        Component 3 & 7.864 & 1.441\\ \hline
    \end{tabular}
\end{table}

For most of the data obtained with all the filters, the profiles were well fitted by this model, although a few were not (we do not use such data in the following discussion).
Fig. \ref{fig:NGC1275_fitting} shows an example of the profile-fitting results.
We integrated each of the emission-model components over 85 $<=$ pixel $<=$ 95 (corresponding to 5.522 arcsec) to obtain a photon count rate. 
Then, we converted the photon count rate to fluxes and corrected them by using the official sensitivity correction function\footnote{http://darts.isas.jaxa.jp/pub/legacy.gsfc.nasa.gov/caldb/data/swift/uvota/bcf/senscorr/} offered by the {\it Swift} team.
Fig. \ref{fig:separation_agn} shows the flux in each band as a function of photon energy (i.e., SED) after correcting for scaling, sensitivity, and Galactic extinction using $E(B$$-$$V) = $0.1399 mag \citep{SF2011ApJ}, assuming ${A_V}/E(B$$-$$V)=$ 3.1 and the extinction law of \citet{Cardelli1989ApJ}.
The values for the Galactic extinction applied here were $A_{V}=0.4433$ mag, $A_{B}=0.5893$ mag, $A_{UUU}=0.6997$ mag, $A_{UVW1}=1.0201$ mag, $A_{UVM2}$=1.3527 mag, and $A_{UVW2}=1.2264$ mag.
As Fig \ref{fig:separation_agn} shows, the AGN emission was brighter at the shorter UV wavelengths than at the longer ones in comparison with the host-galaxy emission.

\clearpage
%
%__________________________________________________________________
\subsection{XRT}

We analyzed the XRT data for NGC 1275 from 2006-01-06 to 2017-04-04. 
The XRT has a photon counting (PC) mode and a windowed timing (WT) mode. 
We performed the data reduction using HEADAS 6.19.

First, we checked for pile-ups in the PC-mode data and found that the data within a 12-arcsec radius from NGC 1275 suffered from pile-up.
To avoid this region, we extracted the spectra from the 12- to 27-arcsec radius from the center of NGC 1275 as the source spectra. 
We extracted the background spectra in the region from the 50- to 60-arcsec radius from NGC 1275.

\begin{figure}[htbp]
  \begin{center}
    \begin{tabular}{c}

      % PC spectrum
      \begin{minipage}{0.4\hsize}
        \begin{center}
        (a)
        \includegraphics[width=6.5cm]{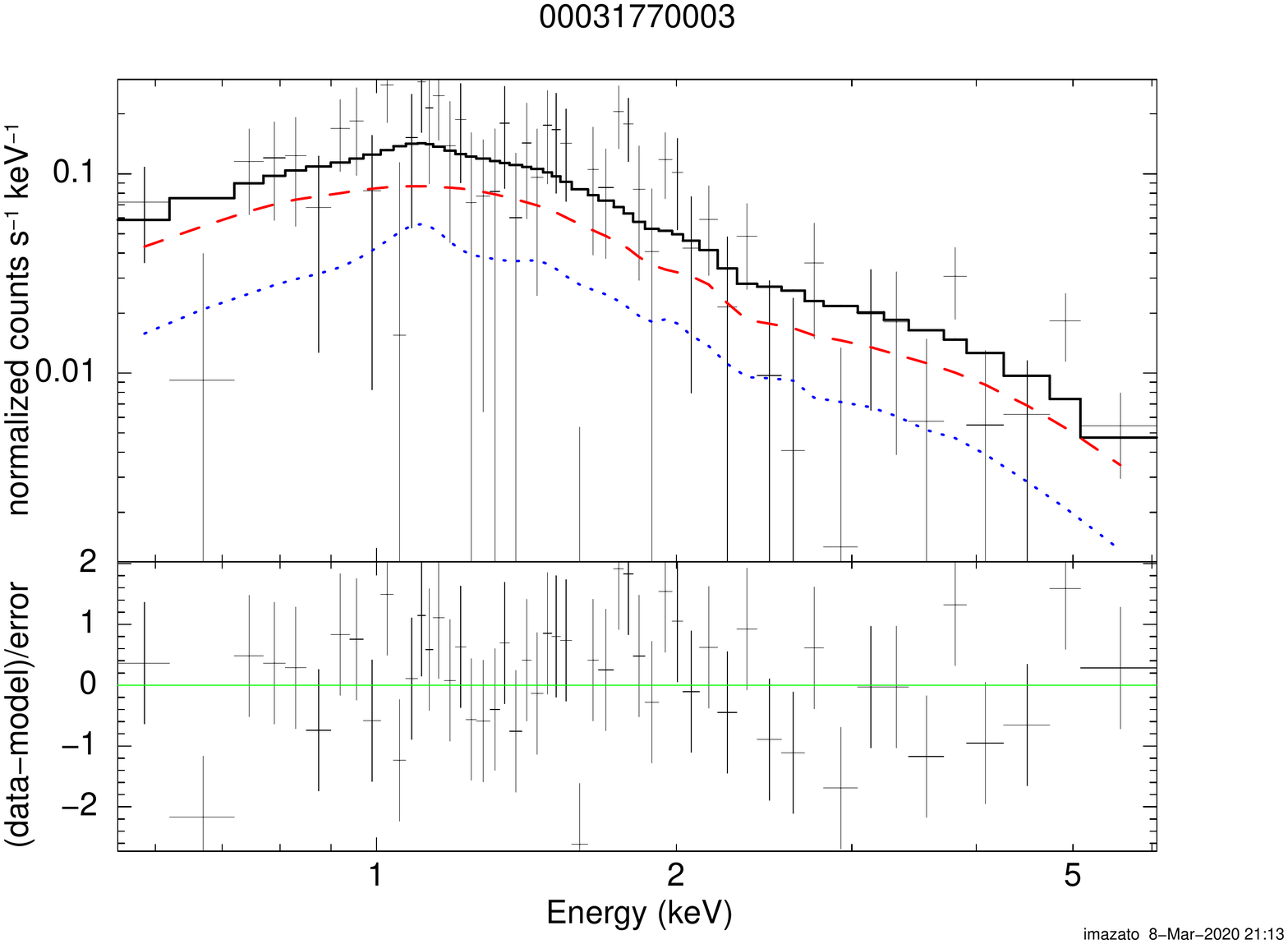}
        \end{center}
      \end{minipage}

      % WT spectrum
      \begin{minipage}{0.4\hsize}
        \begin{center}
        (b)
        \includegraphics[width=6.5cm]{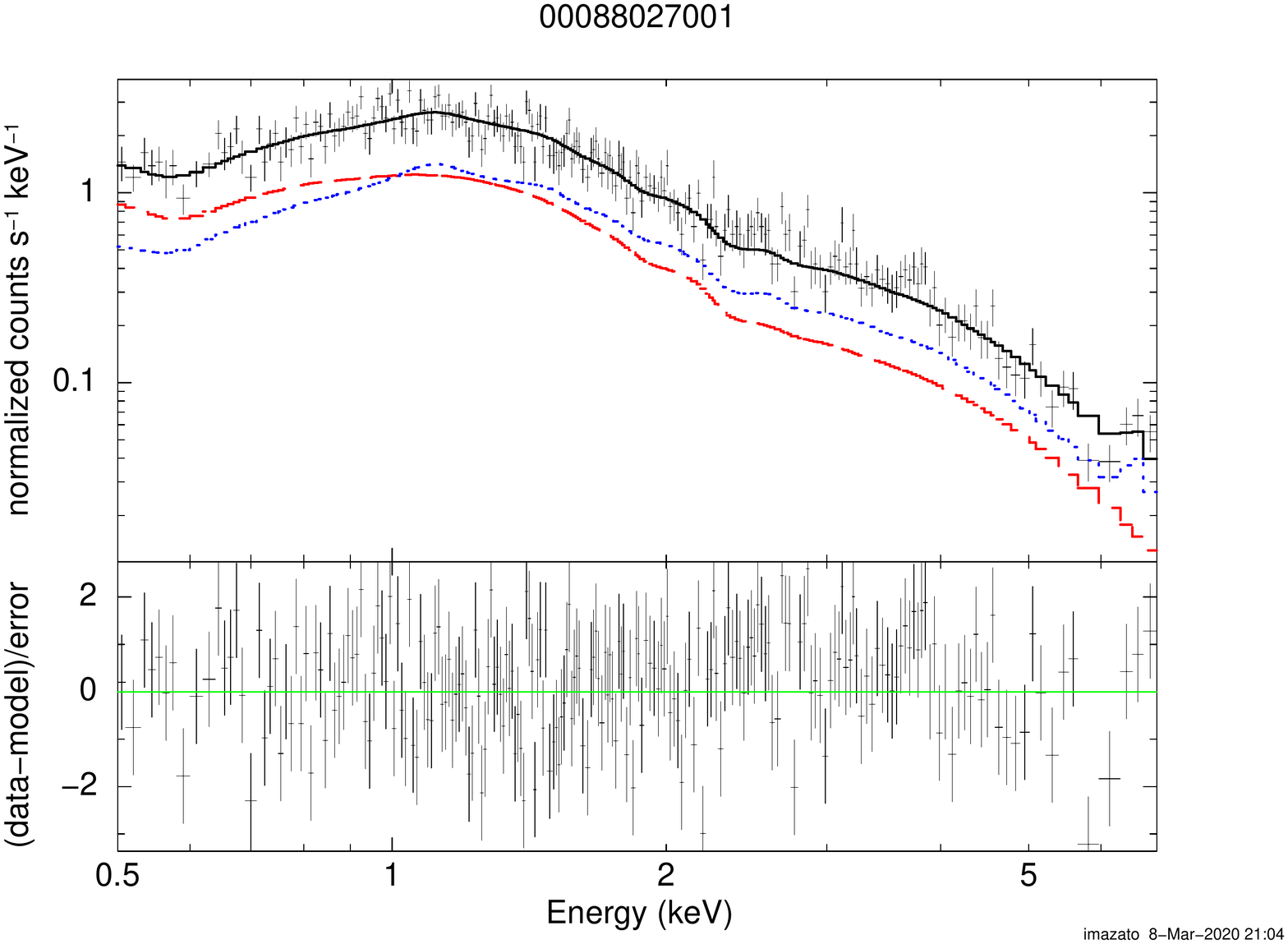}
        \end{center}
      \end{minipage}
      
    \end{tabular}
    \caption{(a) Example of spectral fitting of XRT PC-mode data (obsID = 00031770003, MJD = 55403.55). The red dashed line represents a $pegpwrlw$ model, and the blue dotted line represents an $apec$ model. (b) Example of spectral fitting of XRT WT-mode data (obsID = 00088027001, MJD = 57785.75). The red dashed line represents a $pegpwrlw$ model, and the blue dotted line represents an $apec$ model.}
    \label{fig:spectrum_pc_wt}
  \end{center}
\end{figure}

Second, we used the {\tt xspec} model to fit the PC-mode spectra with a power law plus a plasma model with Galactic absorption: $wabs\times(pegpwrlw + apec)$; $pegpwelw$ is a power-law model for AGN emission. 
We considered cluster emission using the $apec$ plasma model but subtracted the background spectra from the source spectra.
This is because the NGC 1275 cluster emission is not uniform.
We fixed the equivalent hydrogen column density in $wabs$ at 0.138 $\times$ 10$^{22}$ cm$^{-2}$ \citep{Kalberla2005,Hitomi2018PASJ}.
For the $apec$ model parameters, we fixed the plasma temperature at 4 keV, the metal abundances at 0.6 solar, the redshift at 0.01756, and the normalization at 0.00872456.
This normalization value is the median of the parameters obtained by fitting the spectra to all the PC-mode data. 
Fig \ref{fig:spectrum_pc_wt} (a) shows an example of spectral fitting (obsID = 00031770003) to the PC mode data.
% -----
% -------

Analysis of WT-mode data is more complex than that of PC-mode data, because the former have only one-dimensional information and, thus, the cluster emission is integrated along the projection direction in the CCD.
Therefore, the background emission from the cluster becomes relatively stronger than that in the PC-mode, and we must consider background subtraction. 
We describe the analysis in detail bellow.

To estimate the background spectra, including the cluster emission, for the WT-mode data spectra, we used the PC-mode data.
We selected the PC-mode observations where in NGC 1275 was located at the image center and summed these event files.

\begin{figure}
    \centering
    \includegraphics[width=14cm]{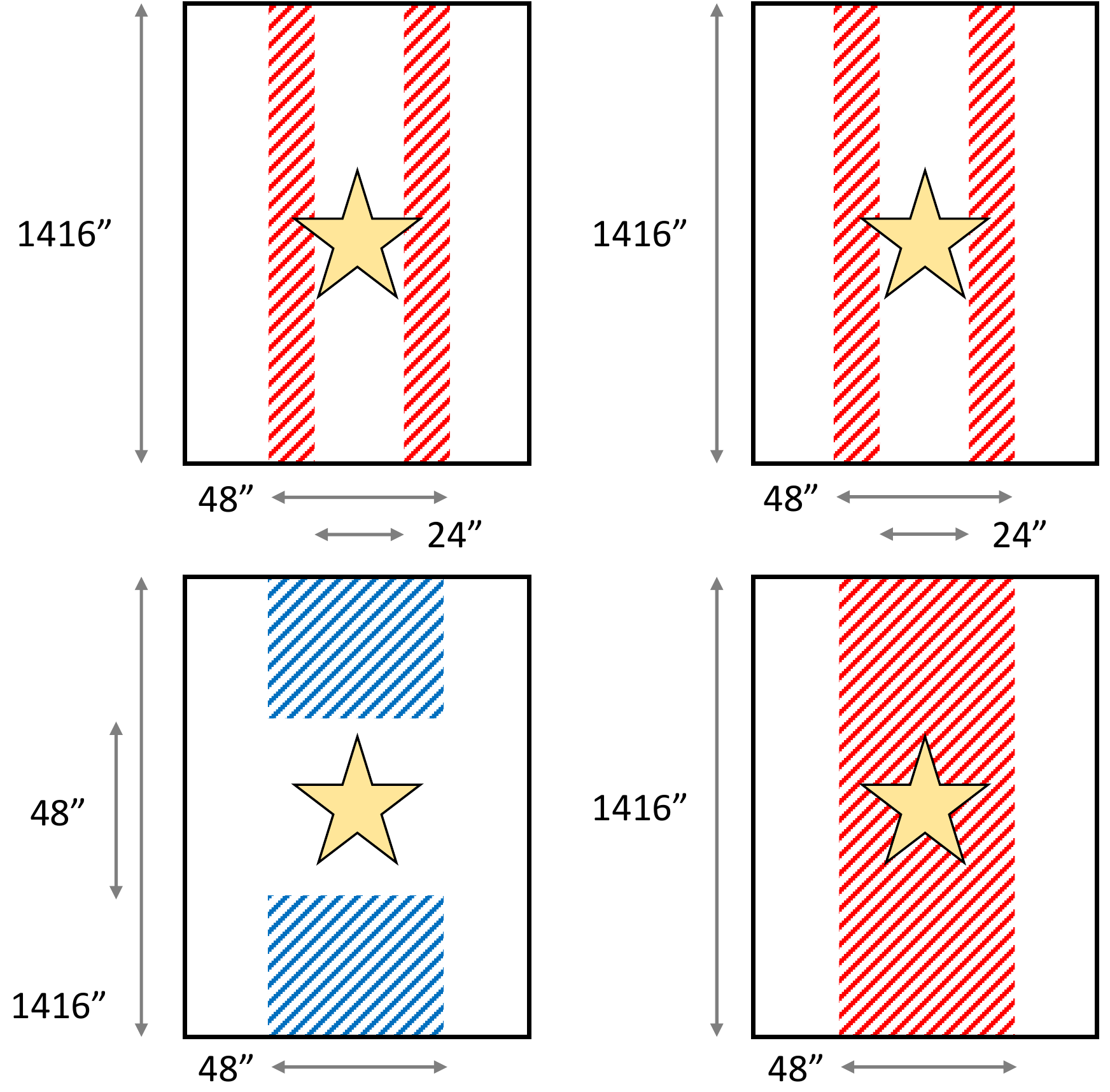}
    \caption{ Integration regions for the spectra for the WT data analysis. 
The top left and right panels show the integration regions for the PC and WT modes, respectively, 
that we used to obtain the arf file correction ratio. The bottom left panel is the integration region for the PC used for estimating the cluster emission in the WT spectra. The bottom right panel is the integration region for the WT data for the AGN analysis.
The yellow star shows the position of NGC 1275.}
    \label{fig:process_of_wt}
\end{figure}

We created spectra and auxiliary response files (arfs).
However, these spectra could not be used for direct subtraction from the WT-mode spectra for the region shown in the bottom right panel of Fig. \ref{fig:process_of_wt} because \citet{Godet2009} reported a systematic difference in the response function in the low energy band.
Thus, our strategy is as follows: First, we represent the background spectra by fitting the PC-mode spectra with the {\tt xspec} model. Second, we modify the arf file for the PC-mode data by considering the systematic differences in the effective areas between the PC and WT models. 
Third, we model the background component using these parameters and the modified arf file to fit the WT-mode spectra.
We fitted the above PC spectra using $wabs \times apec$ and obtained the spectral parameters of the background.

To modify the arf file, we derived the effective area ratio between the PC and WT modes as a function of energy as follows:
We created arf files for each WT and PC observation.
In these calculations, we took the integration region to be that shown as the red areas in the top panels of Fig. \ref{fig:process_of_wt}.
We then added the arf files so-created for all the PC-mode data into one.
Although the region is the same for both the PC and WT modes, the effective areas in these arf files are different. 
Following this, we obtained the ratio of effective areas as a function of energy by dividing the WT arf by the PC arf.

For the AGN analysis, we created spectra and arf files for each set of WT-mode data.
The integration region is shown in the bottom right panel of Fig. \ref{fig:process_of_wt} (corresponding to the spectra extracted from a 48-arcsec width region).
To model the cluster emission, we derived the PC-mode spectrum and arf in the region shown in blue in the bottom left panel of Fig. \ref{fig:process_of_wt} for all the PC data.
We added them into one spectrum or arf, and modified the arf by the ratio obtained above.
Next, we fitted the summed PC spectrum with the $wabs*apec$ model using the modified arf.
Then, we fitted the WT spectra with $wabs \times (pegpwrlw + apec)$, together with the cluster-emission model by using the modified arf and the model parameters for the background region obtained as above.
Fig. \ref{fig:spectrum_pc_wt} (b) shows an example of the resulting spectral fit to the WT-mode data (obsID = 0088027001).

\clearpage
%
%__________________ BAT analysis ____________________________
\subsection{BAT}
\label{chap:BAT}
\begin{figure}
    \centering
    \includegraphics[width=13.7cm]{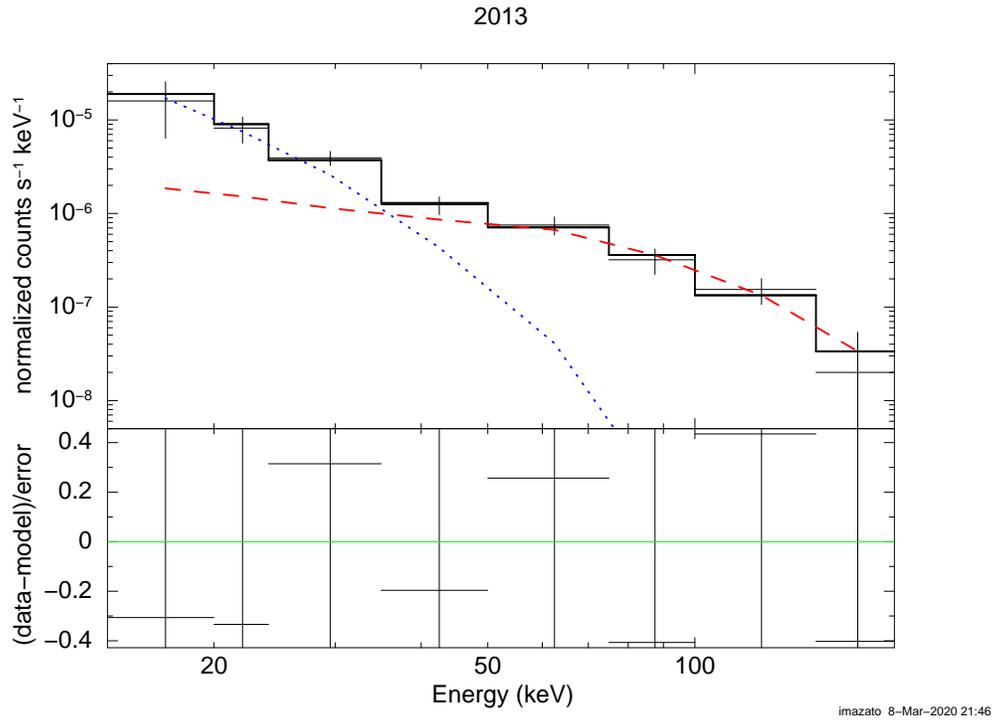}
    \caption{Example of BAT spectral fitting. The black cross data represent the 2013 BAT spectrum. The bold black solid line, red dashed line, and blue dotted line represent the $pegpwrlw + bremss$, $pegpwrlw$, and $bremss$ model, respectively.}
    \label{fig:bat_spectral_fitting_example}
\end{figure}

We extracted the BAT spectrum for NGC 1275 from the 157-month BAT survey catalog (Lien et al. 2020 in prep).
The details of the BAT survey process have been described by \citet{Tueller2010}. 
We re-binned the data using {\tt rebingausslc} to generate the spectral files on a yearly basis and analyzed the BAT data for NGC 1275 from 2005 to 2017.
The BAT spectra were extracted from the 40-arcmin radius region of NGC 1275.
To start, we used the {\tt xspec} model to fit the BAT spectra for NGC 1275 with a power law and a bremsstrahlung model including Galactic absorption: $wabs \times (pegpwrlw + bremss)$.
We fixed the equivalent hydrogen column density in $wabs$ to be 0.138 $\times$ 10$^{22}$ cm$^{-2}$ as well as the Swift/XRT analysis.
However, in most spectral fittings, the errors in the parameters obtained are large.
Therefore, we fixed the parameters of the $bremss$ model at the average value over all the results from the spectral fittings. 
This enabled us to successfully determine the parameters since the cluster emission had to be constant.
Fixing the temperature and normalization of the $bremss$ model at 9.587 keV and 0.0455, respectively, we performed the spectral fittings again to determine the $pegpwrlw$ parameters.
Fig \ref{fig:bat_spectral_fitting_example} shows an example of a spectral fit to the BAT data in 2013.

\clearpage
%
%__________________________________________________________________
\section{Results}
% Lightcurves
\subsection{Light curves}
\begin{figure}
    \centering
    \includegraphics[width=16cm]{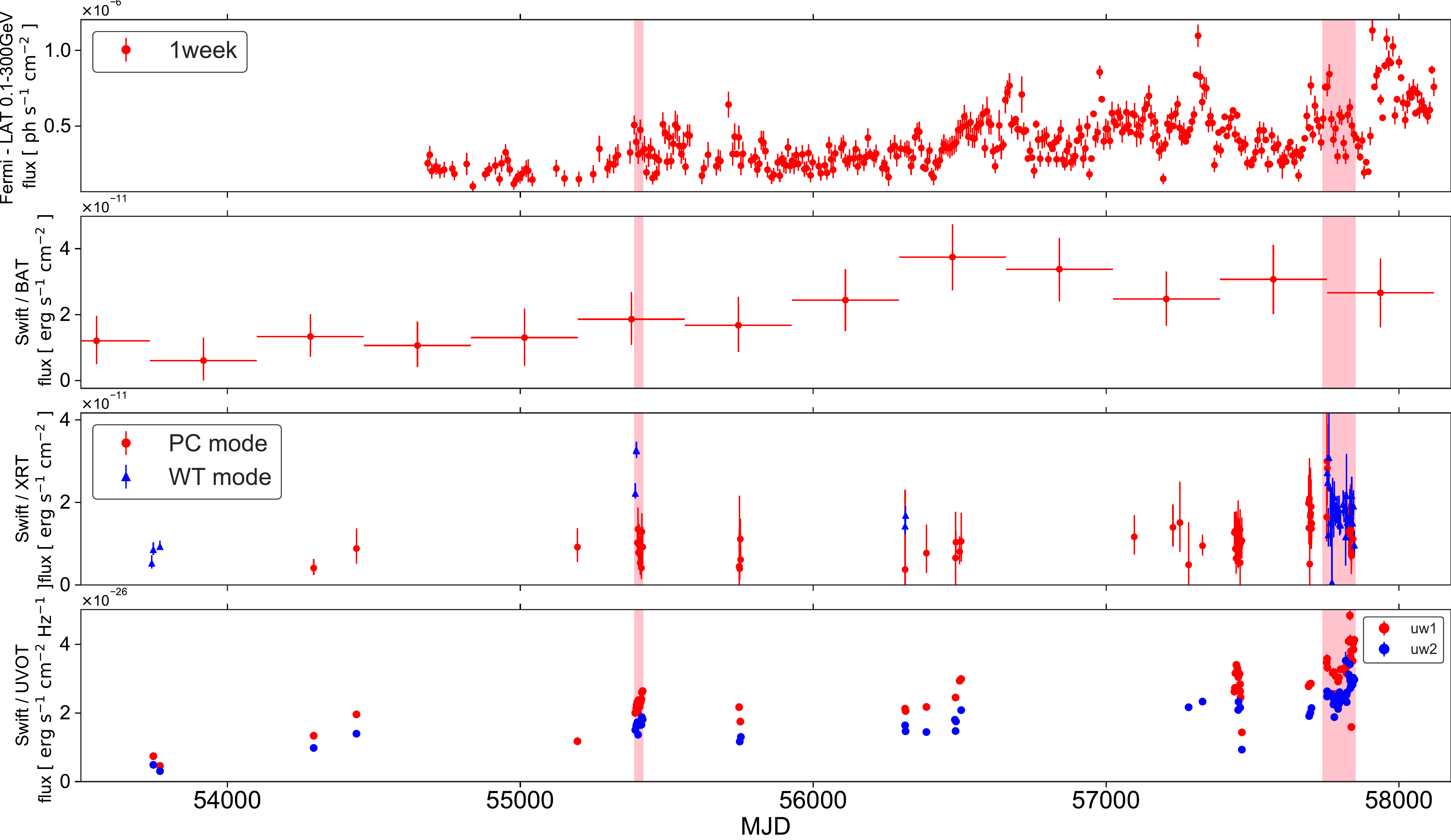}
    \caption{Light curves of NGC 1275. Top: {\it Fermi}/LAT 0.1-300 GeV (MJD $<$ 58119, 2018-01-01); the time bin is one week. Upper middle: {\it Swift}/BAT 50-150 keV; the time bin is one year. Lower middle: {\it Swift}/XRT 5-10 keV; the red and blue circles are the PC- and WT-mode data, respectively. Bottom: {\it Swift}/UVOT; UVW1 (red) and UVW2 (blue).}
    \label{fig:lightcurve_xrt_uvot}
\end{figure}

\begin{figure}
    \centering
    \includegraphics[clip,width=9.cm]{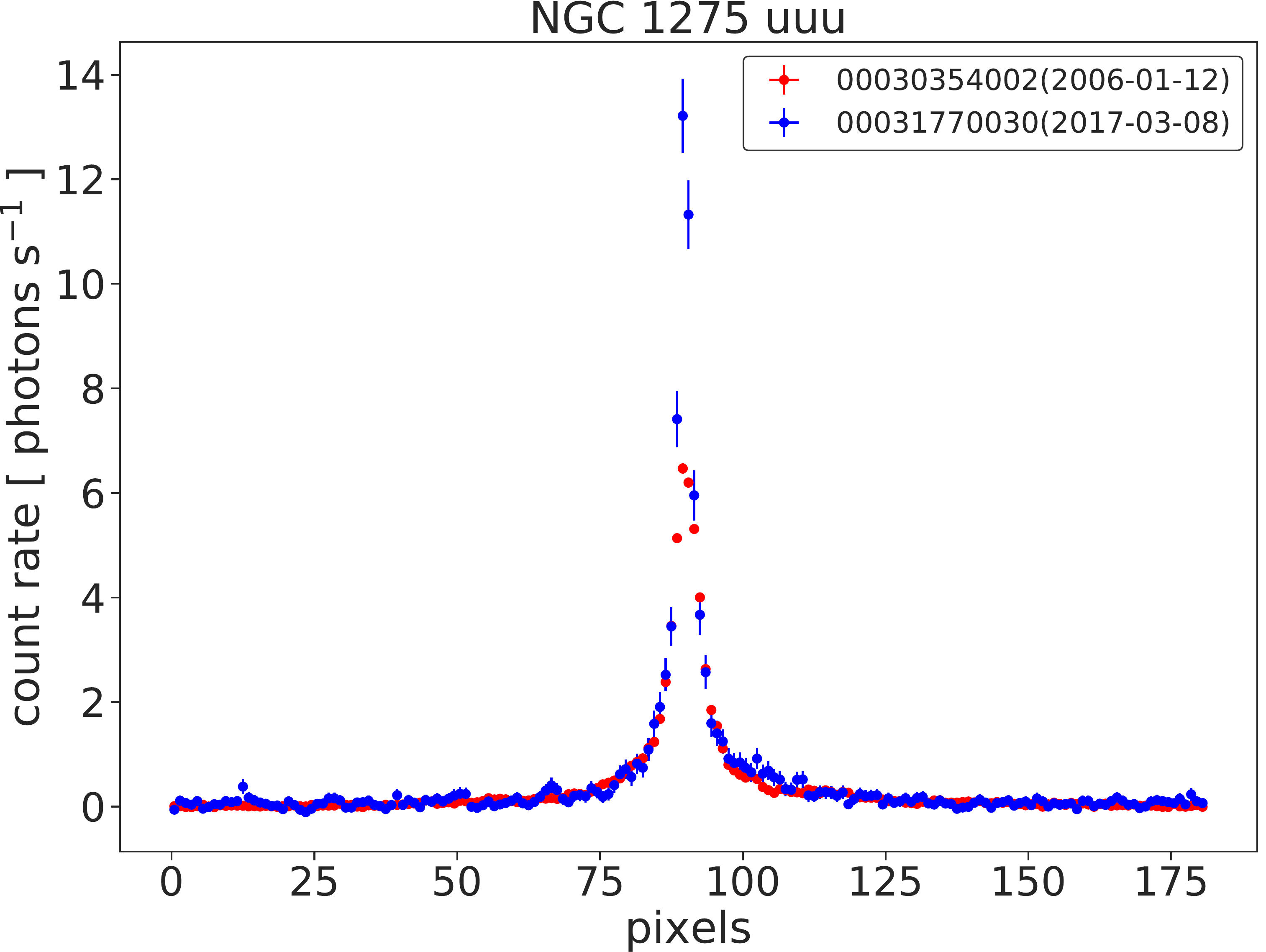}
    \caption{Comparison of the one-dimensional count profiles in the UVOT UUU band between 2006 and 2017. Blue and red represent the bright (2017: MJD = 57820) and faint (2006: MJD = 53747) periods, respectively, in Fig \ref{fig:lightcurve_xrt_uvot}. One pixel is 0.502 arcsec.}
    \label{fig:uvot_comparison}
\end{figure}

\begin{figure}
    \centering
    \includegraphics[clip,width=9.cm]{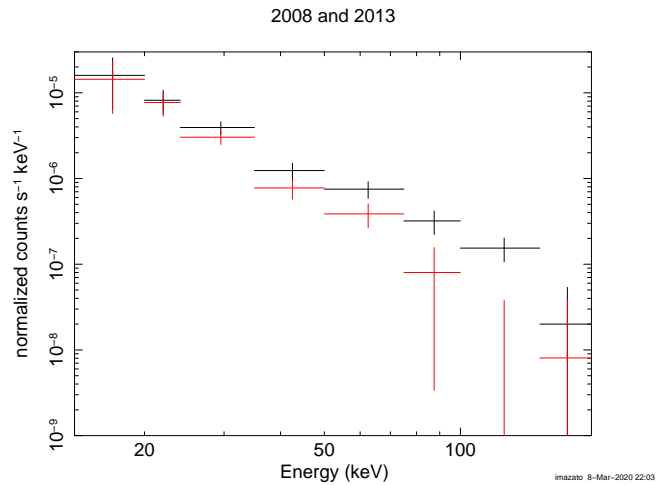}
    \caption{Comparison of the BAT spectra between 2008 and 2013. The black and red crosses represent the bright (2013) and faint (2008) periods, respectively, in Fig \ref{fig:lightcurve_xrt_uvot}.}
    \label{fig:bat_comparison}
\end{figure}

Fig. \ref{fig:lightcurve_xrt_uvot} shows the UV, soft X-ray (5$-$10 keV), hard X-ray (15$-$150 keV), and GeV gamma-ray (0.1$-$300 GeV) light curves of NGC 1275 obtained by {\it Swift}/UVOT, XRT, BAT and {\it Fermi}/LAT.
Both the UV and soft/hard X-ray fluxes obviously increased with the GeV gamma-ray flux. \citet{Fukazawa2018} also reported the soft X-ray flux increase by using {\it Suzaku}/XIS, while UV and hard X-ray flux increases are shown here for the first time.
The flux increase in the UV and hard X-ray band could also be seen in another way: Fig. \ref{fig:uvot_comparison} compares a one-dimensional count profile in the UVOT UUU band 
between the bright (2017) and faint (2006) periods in the UVOT light curve  (Fig. \ref{fig:lightcurve_xrt_uvot}), while Fig. \ref{fig:bat_comparison} compares the BAT spectra between the bright (2013) and faint (2008) periods. 
The difference in the BAT spectra between 2008 and 2013 is less obvious than that in the UVOT profile between 2006 and 2017. 
But the observed flux and photon index of the power-law component in 50$-$150 keV were $10.6_{-6.6}^{+7.3} \times 10^{-12}$ erg cm$^{-2}$ s$^{-1}$ and $1.6_{-1.1}^{+1.1}$ in 2008, and $37.4_{-10.1}^{+10.0} \times 10^{-12}$ erg cm$^{-2}$ s$^{-1}$ and $1.0_{-0.6}^{+0.5}$ in 2013, respectively, indicating that the flux change was significant. 
The confidence range of the fitting results of the BAT spectra was 90\%; the details of the BAT spectral fitting are discussed in  \S\ref{chap:BAT}.
This confirms that the AGN count rate in the UV band and the BAT flux in the higher-energy band actually did increase from 2006 to 2017 and from 2008 to 2013, respectively.

\begin{figure}
    \centering
    \includegraphics[width=16cm]{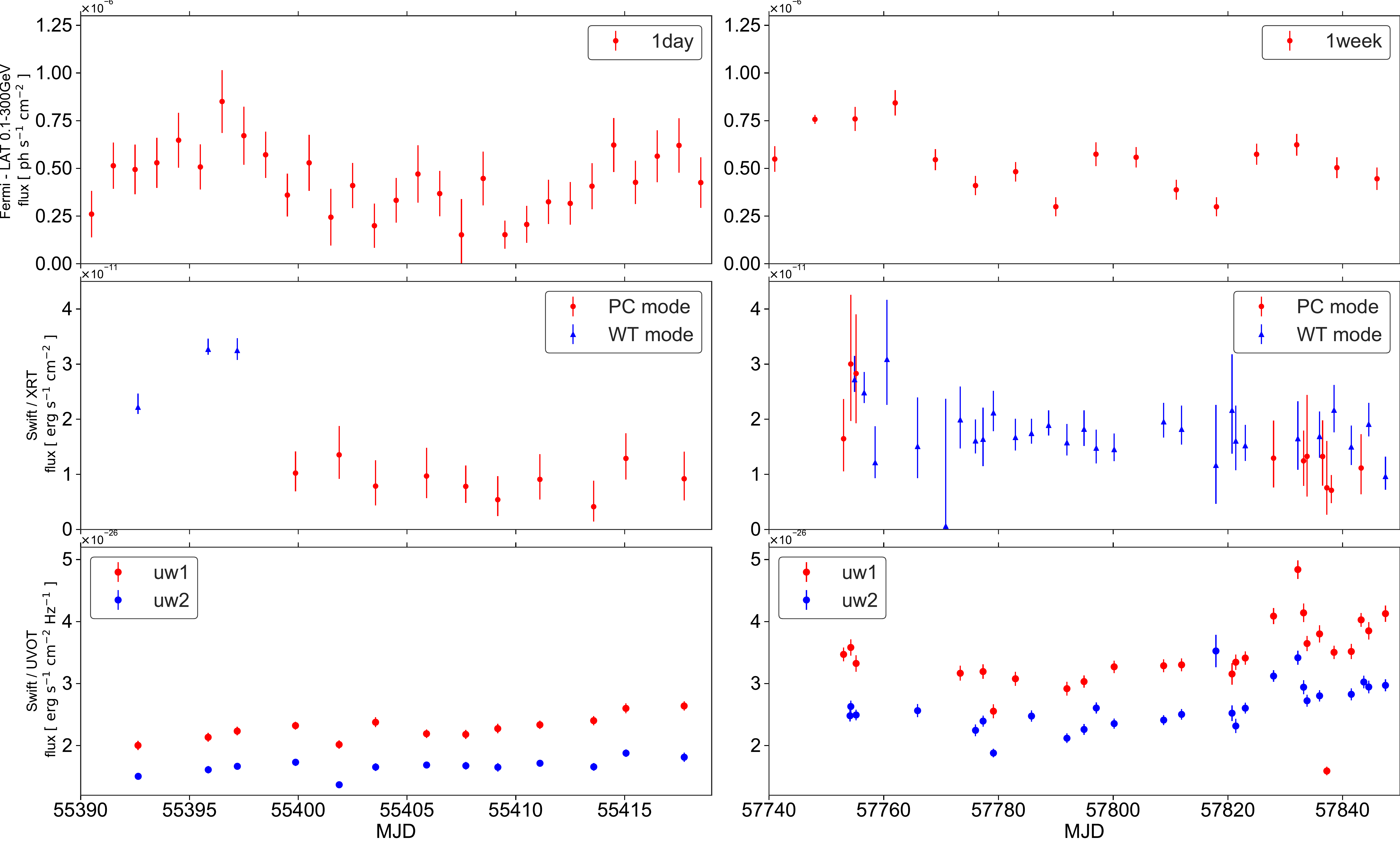}
    \caption{Enlargement of the light curves of NGC 1275 without the BAT data during the periods highlighted by the pink boxes in Fig. \ref{fig:lightcurve_xrt_uvot}. The time bin of {\it Fermi}/LAT is one day.}
    \label{fig:lightcurve_xrt_uvot_short}
\end{figure}

We also observed short-term variability, as in Fig. \ref{fig:lightcurve_xrt_uvot_short}.
Around 55390 $<$ MJD $<$ 55400, a short-term flare in the soft X-ray band appeared, as reported by \citet{Fukazawa2018}; however, flares were not obvious in the UV band.
Around 57820 $<$ MJD $<$ 57840, similar flux variability occurred in the GeV gamma-rays and the UV but was not seen in the soft X-ray band.

\clearpage

% Correlation
\subsection{Correlations among the UV, X-ray, and gamma-ray fluxes}

\begin{figure}[htbp]
  \begin{center}
    \begin{tabular}{c}

      % soft X-ray and GeV gamma-ray
      \begin{minipage}{0.4\hsize}
        \begin{center}
        (a)
          \includegraphics[clip, width=6cm]{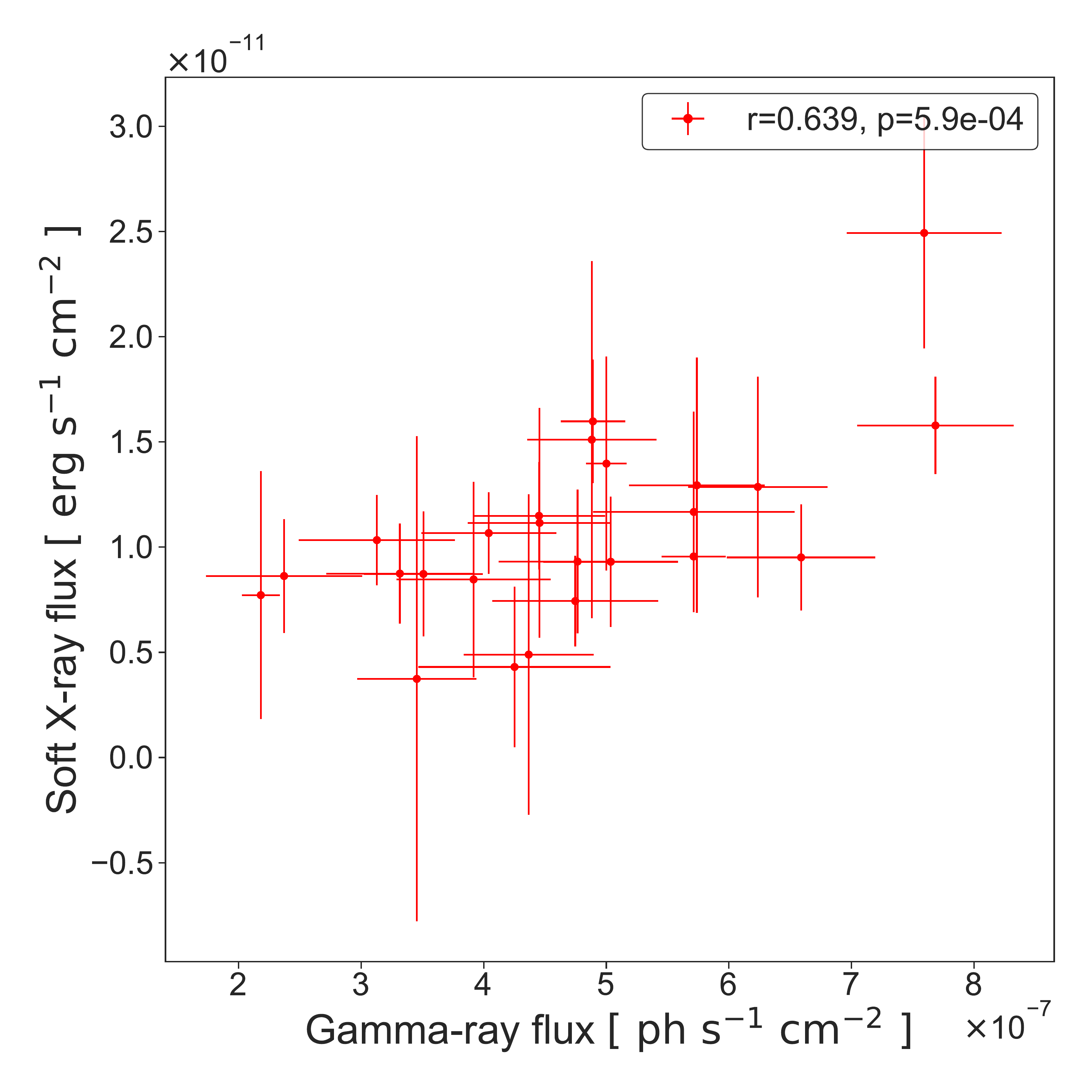}
          \hspace{1.2cm}
        \end{center}
      \end{minipage}

      % hard X-ray and GeV gamma-ray
      \begin{minipage}{0.4\hsize}
        \begin{center}
        (b)
          \includegraphics[clip, width=6cm]{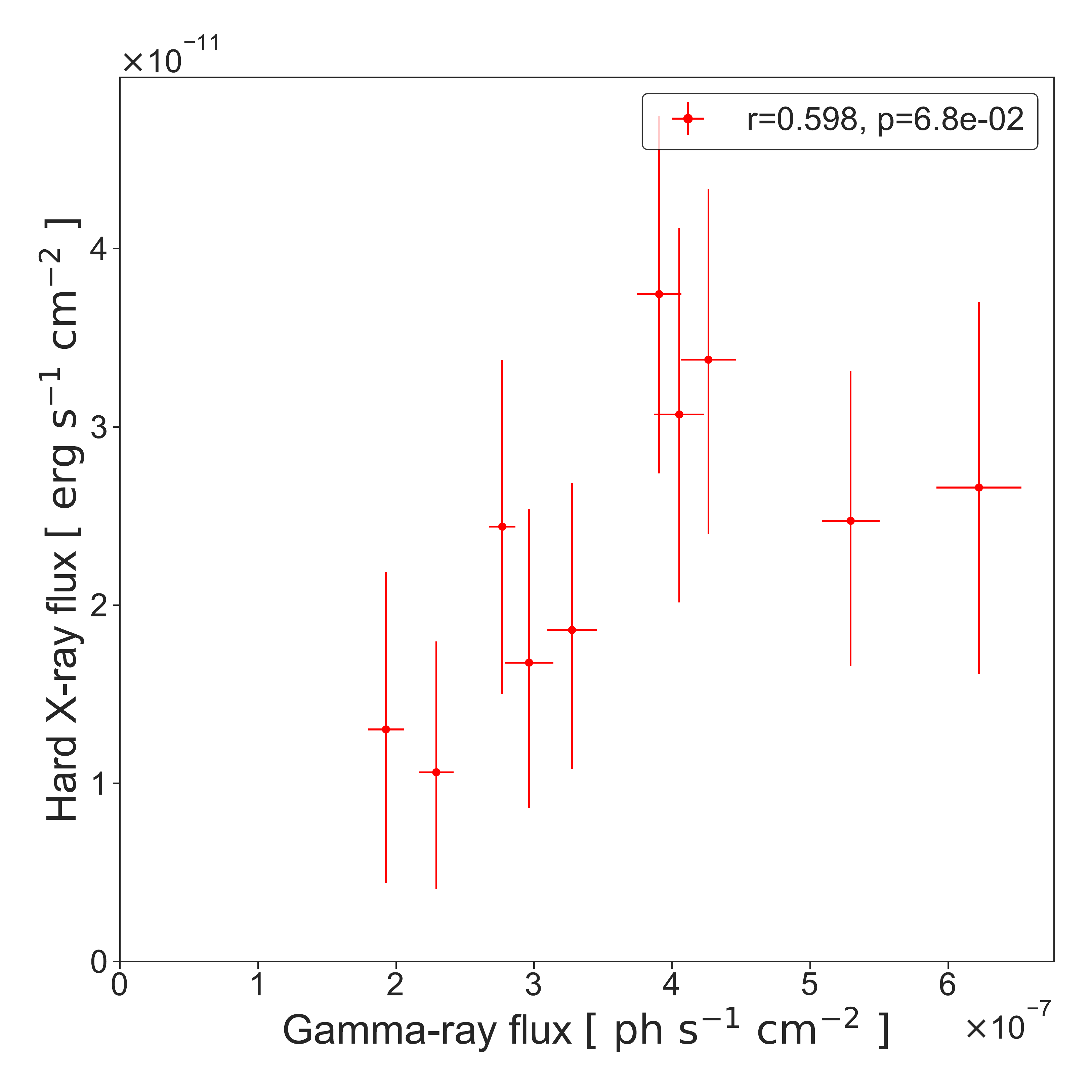}
          \hspace{1.2cm}
        \end{center}
      \end{minipage}\\

      % optical/UV and GeV gamma-ray
      \begin{minipage}{0.4\hsize}
        \begin{center}
        (c)
          \includegraphics[clip, width=6cm]{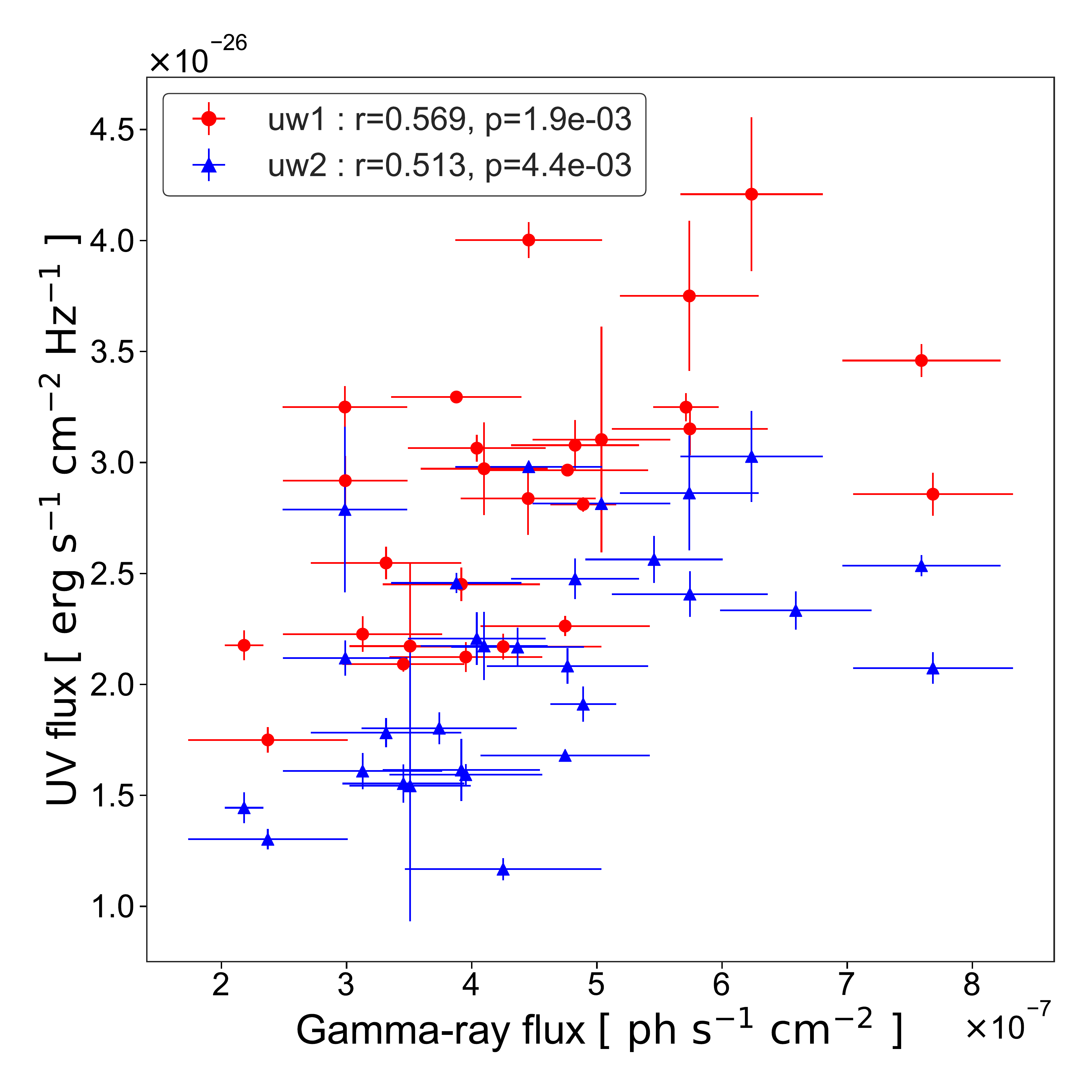}
          \hspace{1.2cm}
        \end{center}
      \end{minipage}

      % optical/UV and soft X-ray
      \begin{minipage}{0.4\hsize}
        \begin{center}
        (d)
          \includegraphics[clip, width=6cm]{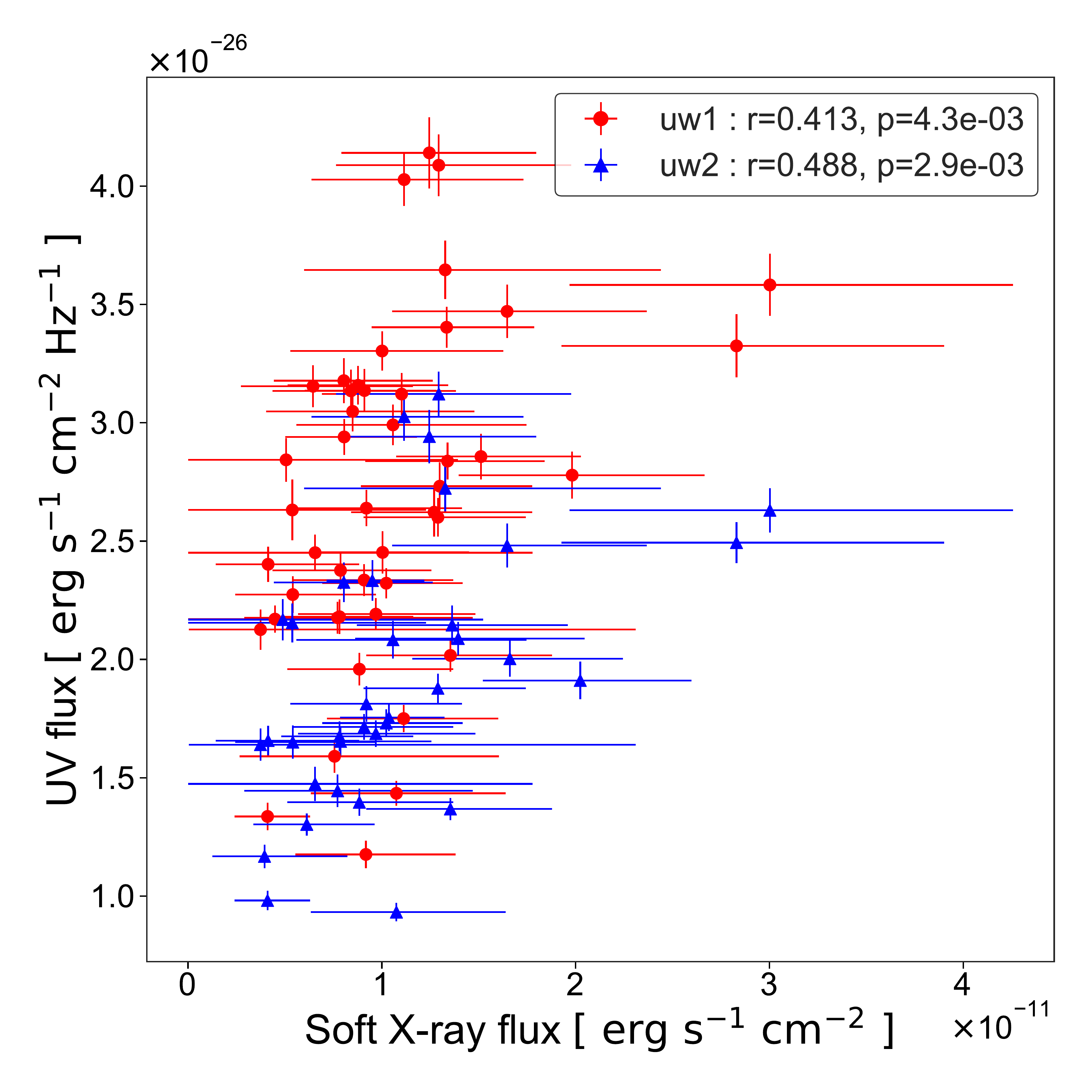}
          \hspace{1.2cm}
        \end{center}
      \end{minipage}

    \end{tabular}
    \caption{(a) Correlation between the soft X-ray (PC-mode data) and GeV gamma-ray fluxes. (b) Correlation between the hard X-rays (BAT) and GeV gamma-rays. (c) Correlation between the UV and GeV gamma-rays. The red and blue circles represent the UVW1 and UVM2 band data, respectively. (d) Correlation between the soft X-rays (PC-mode data) and shorter/longer UV wavelengths. The red and blue circles represent the UVW1 and UVM2 band data, respectively.}
    \label{fig:correlations}
  \end{center}
\end{figure}

To study the origin of the UV and X-ray emission, we investigated the flux correlations of the GeV gamma-rays versus the soft X-rays, the GeV gamma-rays versus the hard X-rays, the GeV gamma-rays versus the UV, and the soft X-rays versus the UV, as shown in Fig. \ref{fig:correlations}. 
Since our {\it Swift}/XRT WT-mode analysis was complicated and, thus, may have had large systematic errors, we considered only the {\it Swift}/XRT PC-mode data as the soft X-ray data in Fig. \ref{fig:correlations}. 
The data in different energy bands were not always from the same time bin. 
In that case, for the data of the energy band with the shorter time bin, we took the average flux over the same time bin period as the other energy band with the longer time bin.
We used Pearson's product-moment correlation coefficient $r_{\rm{pearson}}$, which is widely used to evaluate the linear relationship between two variables in determining the strength of correlations, as well as Spearman's rank correlation coefficient $r_{\rm{spearman}}$, which can also evaluate non-linear relationships.
The Pearson's correlation coefficients are shown in Fig. \ref{fig:correlations}.
To estimate the errors of the correlation coefficients, the data were simulated 10,000 times to regenerate them under the assumption that they were scattered according to a normal distribution; the correlation coefficients were obtained for each of them and the variation was taken as the error.

Fig. \ref{fig:correlations} (a) shows the correlation between the soft X-rays and GeV gamma-ray bands. 
The flux of the GeV gamma-ray data for a one-week time bin is plotted against the average of the flux of the soft X-ray data observed within that week. 
They exhibited a weak positive correlation, with correlation coefficients of $r_{\rm{pearson}}=0.64_{-0.25}^{+0.06}$, $p_{\rm{pearson}}=5.9\times10^{-4}$ and $r_{\rm{spearman}}=0.67_{-0.39}^{+0.03}$, $p_{\rm{spearman}}=2.7\times10^{-4}$.
\citet{Fukazawa2018} reported the Pearson's correlation coefficient of $r_{\rm{pearson}}=$ 0.84 between the {\it Suzaku} soft X-rays and GeV gamma-rays.
Our Pearson's correlation coefficient was smaller than theirs.
Since the {\it Suzaku}/XIS data points in \citet{Fukazawa2018} were evenly sampled over a long period, we considered their data to trace a long-term variability.
However, our {\it Swift}/XRT data were not evenly sampled and, thus, could not trace long-term variations well, resulting in a smaller correlation coefficient.

Fig. \ref{fig:correlations} (b) shows the correlation between the hard X-rays and GeV gamma-ray bands. 
The flux of the hard X-ray data in a one-year time bin is plotted against the average of the flux of the GeV gamma-ray data observed within the same year. 
They exhibited a positive correlation, with correlation coefficients of $r_{\rm{pearson}}=0.60_{-0.27}^{+0.16}$, $p_{\rm{pearson}}=6.8\times10^{-2}$ and $r_{\rm{spearman}}=0.73_{-0.28}^{+0.08}$, $p_{\rm{spearman}}=1.6\times10^{-2}$.
Due to the large p-value ($>1.0\times10^{-2}$), it could not be concluded that the fluxes between the hard X-rays and GeV gamma-rays were correlated.
Even if there was a correlation, this Pearson's correlation coefficient wes smaller than that of the correlation between the soft X-rays ({\it Suzaku}) and GeV gamma-rays \citep{Fukazawa2018}, even though the BAT data points were evenly sampled over a long period.
But when we chose the same period as \citet{Fukazawa2018}, the Pearson's correlation coefficient between the hard X-rays and GeV gamma-rays became more similar, $r_{\rm{pearson}}=0.68_{-0.24}^{+0.14}$. 
However, the p-values were still large, i.e., $p_{\rm{pearson}}=6.2\times10^{-2} > 1.0\times10^{-2}$, which means that they could not be considered correlated.

Fig. \ref{fig:correlations} (c) also shows the correlations between the UV and GeV gamma-ray bands, with correlation coefficients of $r_{\rm{pearson}}=0.57_{-0.21}^{+0.07}$, $p_{\rm{pearson}}=1.9\times10^{-3}$ and $r_{\rm{spearman}}=0.56_{-0.21}^{+0.12}$, $p_{\rm{spearman}}=2.5\times10^{-3}$ (UVW1 and GeV gamma-ray bands) and $r_{\rm{pearson}}=0.51_{-0.22}^{+0.09}$, $p_{\rm{pearson}}=4.4\times10^{-3}$ and $r_{\rm{spearman}}=0.57_{-0.23}^{+0.10}$, $p_{\rm{spearman}}=1.4\times10^{-3}$ (UVW2 and GeV gamma-ray bands). 
The flux of the GeV gamma-ray data for a one-week time bin is plotted against the average of the flux of the UV data observed within the same week in Fig. \ref{fig:correlations} (c). 
These correlations were somewhat smaller than those of the soft X-ray and GeV gamma-ray bands, indicating that the dominant emission was not from the jet, whose emission usually shows a flux correlation between optical and GeV gamma-rays \citep{Itoh2016}. 
However, we noted a similar long-term flux increase of the UV band with the GeV gamma-ray flux.

Figure 10 (d) plots the fluxes of the XRT and UVOT for the same observation ID. 
The correlation coefficients were $r_{\rm{pearson}}=0.41_{-0.35}^{+0.06}$, $p_{\rm{pearson}}=4.3\times10^{-3}$ and $r_{\rm{spearman}}=0.44_{-0.42}^{+0.06}$, $p_{\rm{spearman}}=2.2\times10^{-3}$ (UVW1 and soft X-ray bands) and $r_{\rm{pearson}}=0.49_{-0.37}^{+0.06}$, $p_{\rm{pearson}}=2.9\times10^{-3}$ and $r_{\rm{spearman}}=0.56_{-0.46}^{+0.04}$, $p_{\rm{spearman}}=4.1\times10^{-4}$ (UVW2 and soft X-ray bands).
There was no correlation, or, at most, a very weak one, between the UV and soft X-ray fluxes, indicating that they varied independently, with only a small contribution from the common jet component.

\begin{figure}[htbp]
  \begin{center}
    \begin{tabular}{c}

      % SED UVOT to BAT
      \begin{minipage}{0.4\hsize}
        \begin{center}
        (a)
          \includegraphics[clip,width=6.5cm]{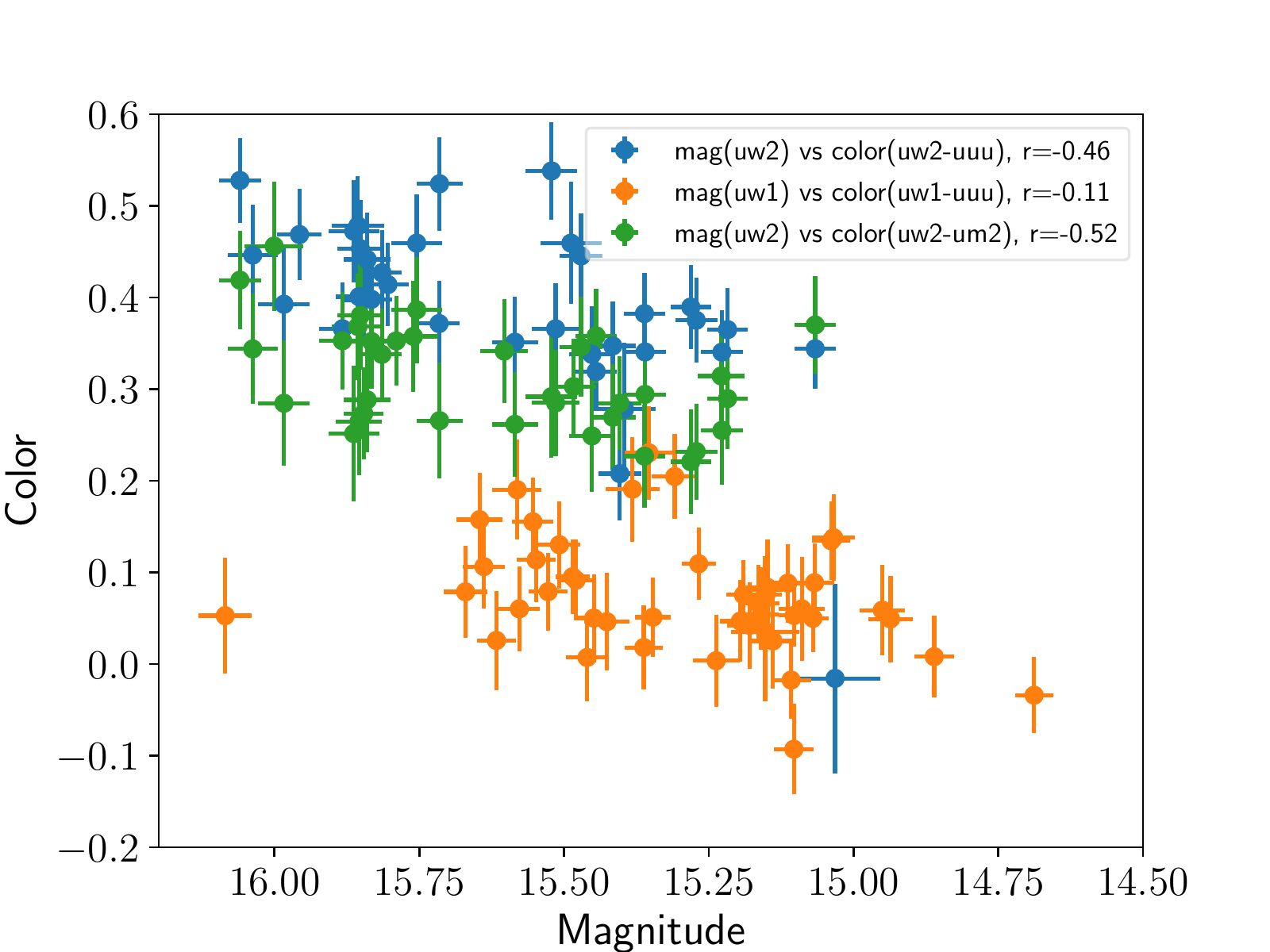}
        %   \hspace{1.2cm}
        \end{center}
      \end{minipage}

      % SED UVOT with BB fit
      \begin{minipage}{0.4\hsize}
        \begin{center}
        (b)
          \includegraphics[clip,width=6.5cm]{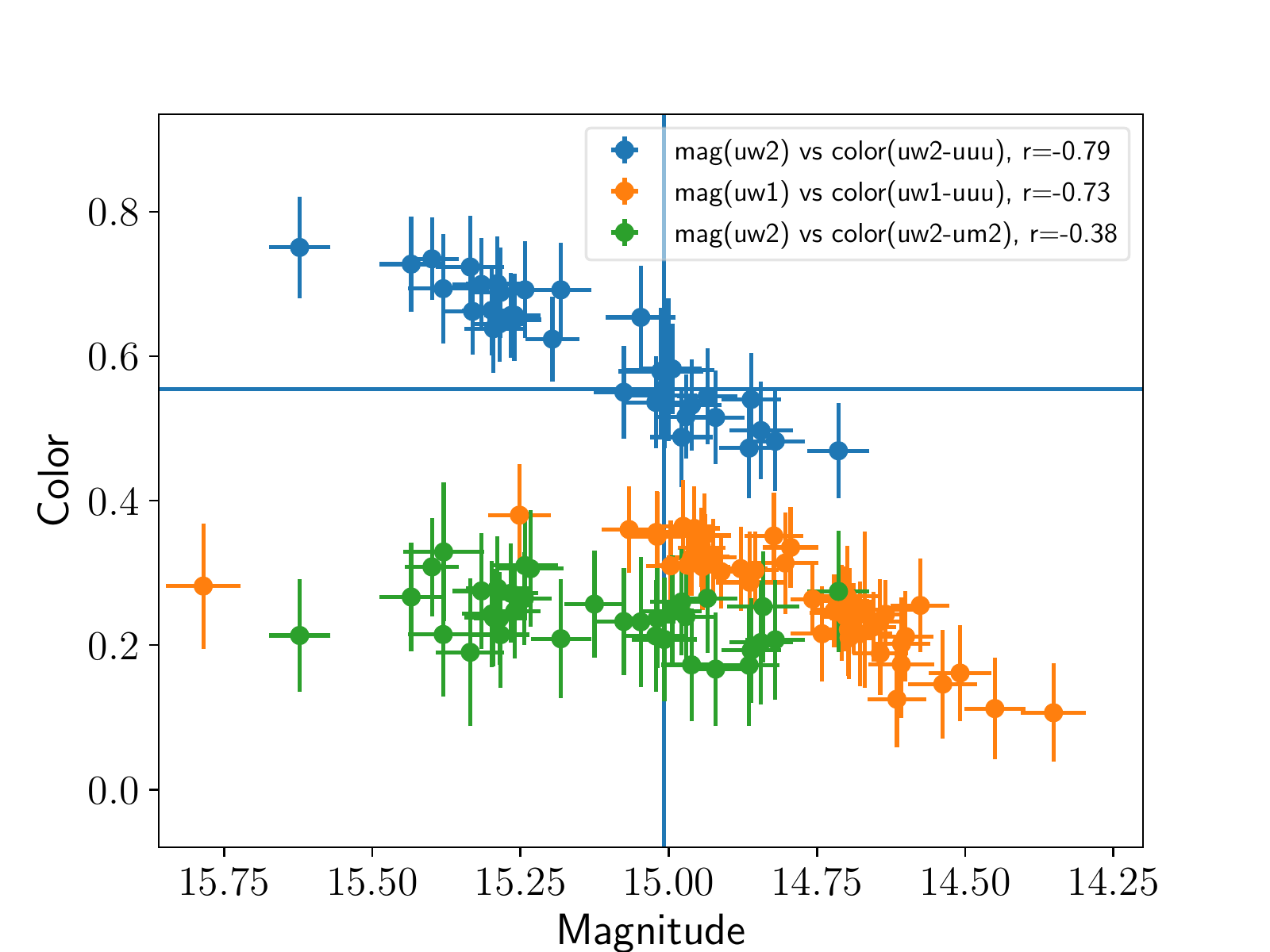}
        %   \hspace{1.2cm}
        \end{center}
      \end{minipage}
    \end{tabular}
    \caption{(a) Color-magnitude diagram of the AGN component of NGC 1275. The blue, orange, and green circles represent the UVW2 magnitude vs. UVW2$-$UUU color, the UVW1 magnitude vs. UVW1$-$UUU color, and the UVW2 magnitude vs. UVW2$-$UVM2 color, respectively. (b) Color-magnitude diagram of the sum of the AGN and host-galaxy components of NGC 1275.}
    \label{fig:color_mag}
  \end{center}
\end{figure}

% Color-Magnitude diagram
% \subsection{Color-Magnitude diagram}
Fig. \ref{fig:color_mag} (a) shows the UV color-magnitude diagram without the host-galaxy component. 
% There are no obvious feature in the UV bands, and the color is almost constant, or weak bluer-when-brighter trend. 
The correlation coefficients were $r_{\rm{pearson}}=-0.11$, $p_{\rm{pearson}}=4.8\times10^{-1}$ (UVW1 and UUU bands) and $r_{\rm{pearson}}=-0.52$, $p_{\rm{pearson}}=6.6\times10^{-4}$ (UVW2 and UVM2 bands).
This result means that the UV radiation of the AGN component at short wavelengths was characterized by a stronger negative correlation than at long wavelengths.
Note that the p-value for the long-wavelength side (UVW1 and UUU bands) was above 0.01, which means that it could not be concluded that there was a correlation.

% This suggests that the UV emission is not dominated by the jet component.

\clearpage
%

% SED
\subsection{SED in the UV and X-ray bands}

\begin{figure}[htbp]
  \begin{center}
      % SED UVOT to BAT
      \begin{minipage}[b]{0.4\hsize}
        \begin{center}
        (a)
          \includegraphics[clip,width=6.5cm]{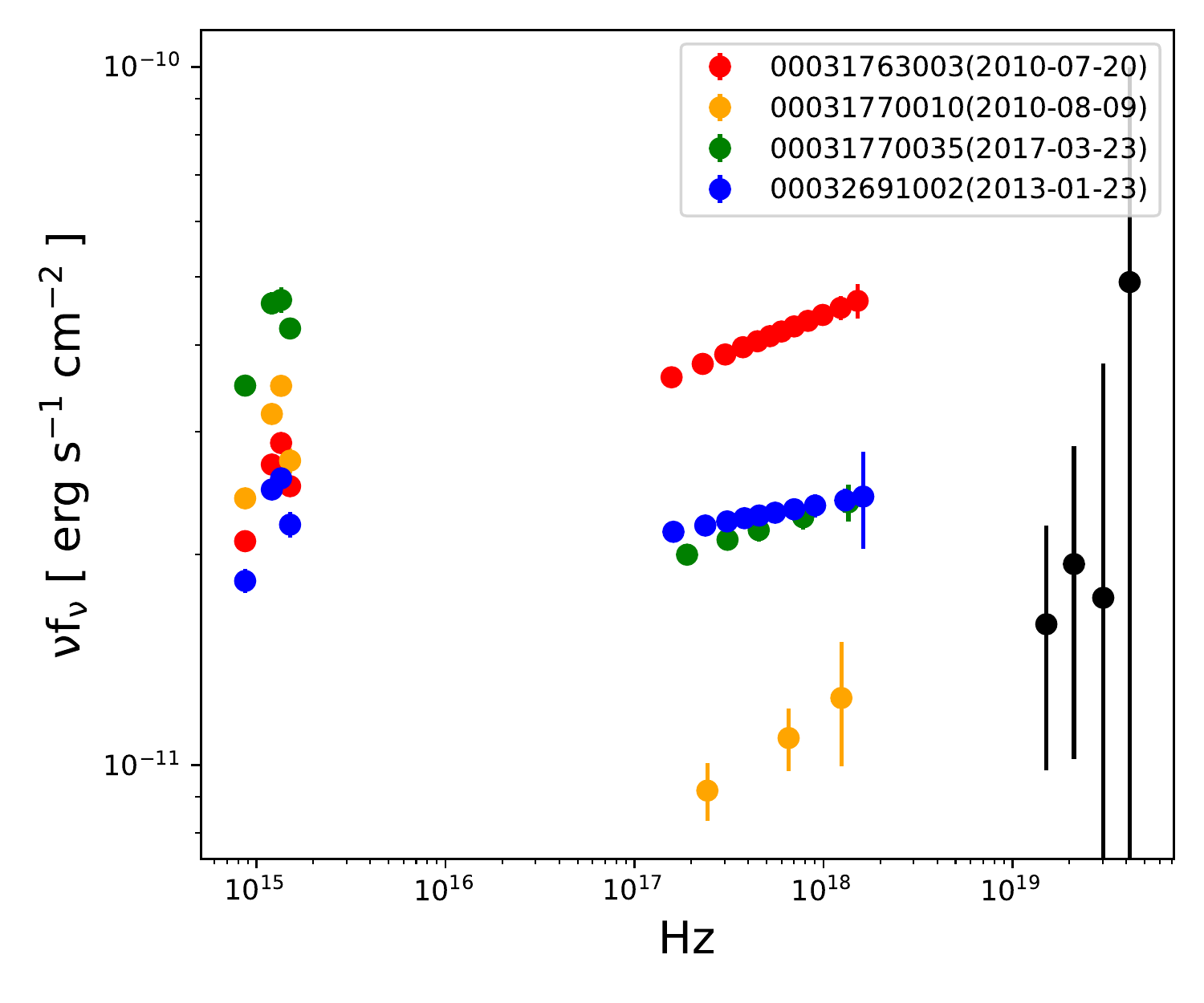}
        %   \hspace{1.2cm}
        \end{center}
      \end{minipage}
      % SED UVOT with BB fit
      \begin{minipage}[b]{0.4\hsize}
        \begin{center}
        (b)
          \includegraphics[clip,width=6.5cm]{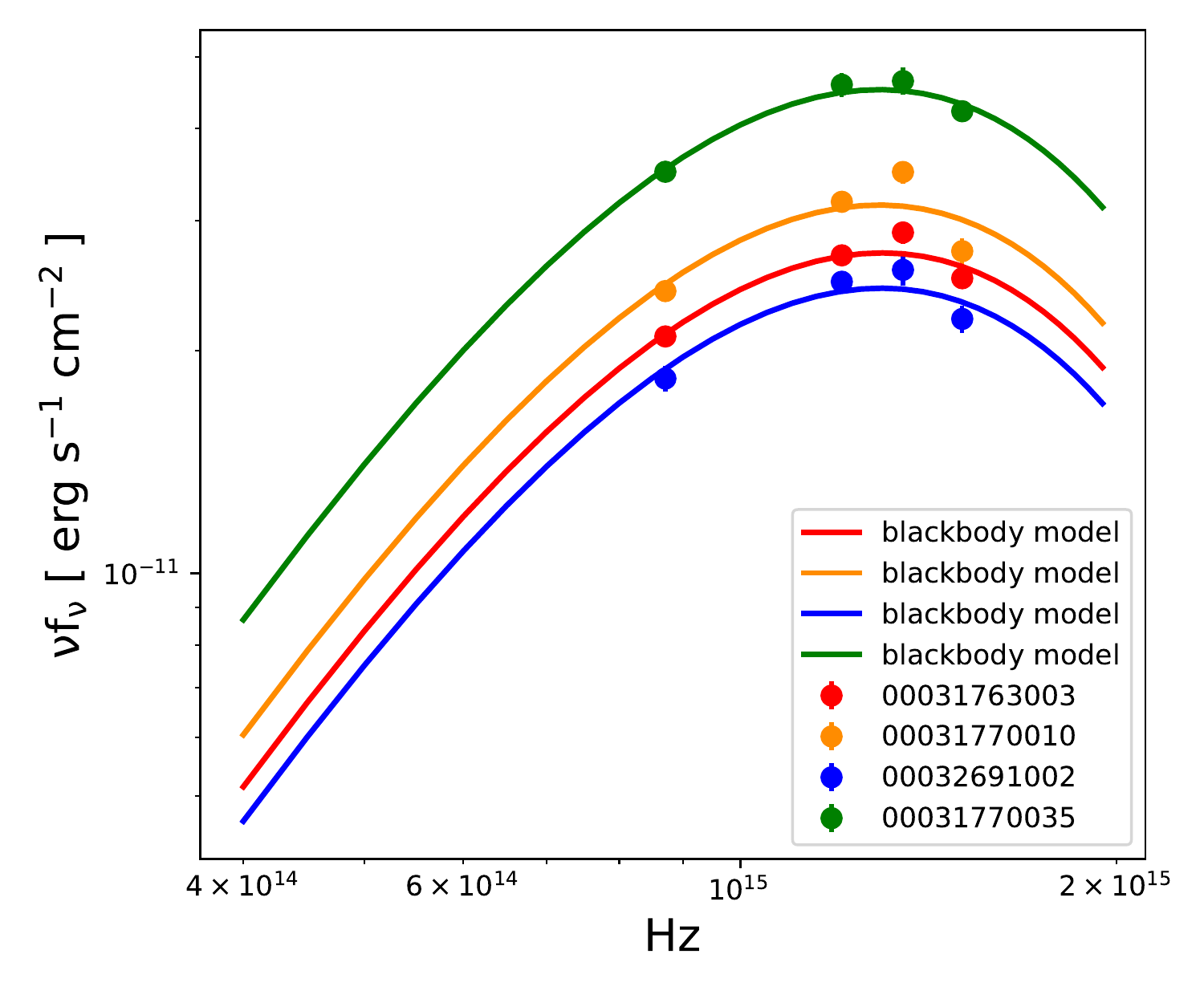}
        %   \hspace{1.2cm}
        \end{center}
      \end{minipage} 
      
    \caption{(a) UV to X-ray SED of NGC 1275. The colors represent the obsIDs. Red is 00031763003 (2010/7/20, MJD = 55397; X-ray flare period); yellow is 00031770010 (2010/8/9, MJD = 55417; X-ray low flux state); green is 00031770035 (2017/3/23, MJD = 57835; UV high flux state); blue is 00032691002 (2013/1/23, MJD = 56315; UV low flux state); black is the average and the standard deviation of the flux over all periods of BAT data. To remove the component due to the Perseus cluster, we show the BAT data only above 50 keV. (b) The SED in the UV band. The solid lines represent the best-fit single-temperature blackbody radiation model.}
    \label{sfig:SED}
  \end{center}
\end{figure}

In this subsection, to understand the UV to X-ray spectral shape, Fig. \ref{sfig:SED} (a) shows examples of the SEDs of NGC 1275 in four periods.
The first spectrum was derived from the X-ray flare at MJD = 55397 (obsID = 00031763003, July 20, 2010).
The second one was derived from the X-ray low flux state at MJD = 55417 (obsID = 00031770010, August 9, 2010).
The third one was from the UV high flux state at MJD = 57835 (obsID = 00031770035, March 23, 2017).
The last one was from the UV low flux state at MJD = 56315 (obsID = 00032691002, January 23, 2013).
% Almost all the SEDs in the X-ray band can be smoothly connected from soft X-rays (XRT) to hard X-rays (BAT) with a single component. 
% As discussed later, this is a combination of disk/corona emission and jet emission, both of which have a similar spectral shape in the X-ray band \textbf{\textcolor{red}{\citep{Fukazawa2011}}}.
All UV SEDs showed a narrow spectral shape with a peak in the UV band. 
The UV and X-ray spectral shape could not be reproduced by non-thermal jet emission with the electron energy distribution assumed to have a broken power-law shape; evidently, only the UV spectra could be represented by a single-temperature blackbody curve, as shown by the solid lines in Fig. \ref{sfig:SED}(b).
The obtained temperature for the single-temperature blackbody model was around 15000$-$18000 K for all the UVOT observational data.
Table \ref{tab:fitting_res_SED} shows information on the X-ray spectra and the best-fit parameters of the UV blackbody model for the four periods in Fig. \ref{sfig:SED}.

% -------

\begin{table}[hbtp]
    \caption{Fitting parameters for the X-ray and UV fluxes in Fig. \ref{sfig:SED} (b).}
    \label{tab:fitting_res_SED}
    \centering  
    \begin{tabular}[t]{lcccc}
        \hline \hline
        ObsID & 00031763003 & 00031770010 & 00031770035 & 00032691002 \\ \hline
        soft X-ray power-law flux & 32.5$^{+2.2}_{-1.8}$ & 9.2$^{+4.9}_{-3.9}$ & 16.9$^{+4.5}_{-3.9}$ & 16.8$^{+2.2}_{-2.0}$ \\
        soft X-ray power-law photon index & 1.89$^{+0.04}_{-0.04}$ & 1.83$^{+0.33}_{-0.30}$ & 1.91$^{+0.18}_{-0.19}$ & 1.95$^{+0.08}_{-0.09}$ \\
        UV blackbody temperature & 1.59$\pm$0.05 & 1.6$\pm$0.1 & 1.59$\pm$0.07 & 1.58$\pm$0.03 \\
        % UV blackbody radius & 7.6$\pm$0.5 & 8.3$\pm$1.0 & 7.2$\pm$0.6 & 10.0$\pm$0.4 \\ \hline
        UV blackbody radius & 3.1$\pm$0.2 & 3.4$\pm$0.5 & 4.0$\pm$0.2 & 3.0$\pm$0.3 \\ \hline
    \end{tabular}
    The units for each value in the table are 10$^{-12}$ erg/cm$^2$/s and 10$^4$ K for the soft X-ray flux and UV black body temperature, in that order.
    UV blackbody radius is in Schwarzschild radius units.
    The Schwarzschild radius is 2.36$\times$10$^{14}$ cm, assuming the mass of NGC 1275 is M=8$\times$10$^{8}$M$_\odot$ \citep{Scharw2013}.
    
\end{table}

% -------
\clearpage

%__________________________________________________________________
\section{Discussion}

% optical/UV origin discussion   
\subsection{Origin of the UV emission}
\label{optuv_origin}
We found that the UV flux of the nucleus of NGC 1275 increased from 2008 to 2017 along with the X-ray and GeV gamma-ray fluxes.
This suggests that jet emission contributes to the UV emission from NGC 1275 at a certain level (Figs. \ref{fig:lightcurve_xrt_uvot} and \ref{fig:lightcurve_xrt_uvot_short}); however more research is needed. 
In fact, there were periods in which the UV and GeV gamma-rays varied independently from each other on a short timescale.
The UV band lay around the peak of synchrotron emission from the jet; thus, it had no unique features, something like very flat. 
Therefore, the jet contribution could not be concluded from the spectral studies.

Our result for the UV color-magnitude diagram (Fig. \ref{fig:color_mag} (b)) was different from that of the optical (I, R, and V) bands shown by \citet{Yuan2015}. 
From the complicated spectra on the I and R bands and the clear bluer-when-brighter trend on the V band, which was similar to many BL Lacs \citep{Ikejiri2011}, they suggested that jet emission makes a contribution at the V band. 
However, they compared the relations between the spectral-index and flux in the three optical bands without separating the AGN emission from the host-galaxy emission of NGC 1275. 
% When we created color-magnitude diagram in their way (Fig.\ref{fig:color_mag} (b)), we found that the bluer-when-brighter trend was more pronounced in the UVW2-UUU bands and UVW1-UUU bands than in Fig. \ref{fig:color_mag} (a). 
When we created a color-magnitude diagram including the host-galaxy component as they did (Fig.\ref{fig:color_mag} (b)), we found that the bluer-when-brighter trend on the longer UV wavelength (UVW1 and UUU bands) was pronounced ($r_{\rm{pearson}}=-0.73$, $p_{\rm{pearson}}=4.5\times10^{-9}$); this was the same trend as the V band of \citep{Yuan2015} rather than ($r_{\rm{pearson}}=-0.11$, $p_{\rm{pearson}}=4.8\times10^{-1} > 0.01$) of the AGN component alone.
% This occurs because there is a brighter galaxy component of NGC 1275 at the longer wavelength. 
However, on the shorter UV wavelength side (UVW2 and UVM2 bands), the bluer-when-brighter trend was weaker or nonexistent including the host-galaxy component ($r_{\rm{pearson}}=-0.38$, $p_{\rm{pearson}}=1.6\times10^{-2} > 0.01$) than that ($r_{\rm{pearson}}=-0.52$, $p_{\rm{pearson}}=6.6\times10^{-4}$) of the AGN component alone.

% However, the color-magnitude diagram of the UVW2-UVM2 bands shows that color is constant over one magnitude or is only slightly bluer-when-brighter. 
The difference between the optical (V band) and shorter UV wavelengths (UVW2-UVM2) in the color-magnitude diagram including the host-galaxy (Fig. \ref{fig:color_mag} (b)) implied that the different AGN emission was dominant in the optical and UV wavelengths.
Since \citep{Yuan2015} pointed out that the emission of the jet may be dominant in the optical wavelength (V band), the emission of the accretion disk may be dominant in the UV wavelengths (UVW2-UVM2).
However, it should be noted that the difference between this study and Yuan et al.'s 2015 may be due to the insufficient statistics of our UV data.

% This difference between the R-band \citep{Yuan2015} and the our shorter UV wavelength (UVW2-UVM2 bands) also means that the origin of the UV emission in our data is dominated by another component and not by the jet.
% Emission from the accretion disk has been suggested to be more dominant at the shorter UV wavelengths than at the longer ones.
% However, another reason for this difference between the optical and UV bands may be due to insufficient statistics of our UV data. 

%\clearpage

% Accretion disk parameters discussion

The UV spectra of the NGC 1275 nucleus had a narrow peaked spectral shape (Fig.\ref{sfig:SED} (b)). 
This could be represented by a single-temperature blackbody, which was an approximation of the standard disk \citep{Shakura1973}.
The standard disk model predicts multi-temperature blackbody emission, with a flatter spectral shape than the single-temperature blackbody in the lower-energy band.
However, the observed UV spectra did not match a multi-temperature blackbody.
Nevertheless, to examine whether the blackbody parameters obtained from the spectral fitting were reasonable or not, the radius of the emission region of the accretion disk in units of the Schwarzschild radius for a non-rotating black hole was derived using the single-temperature blackbody model with Stefan–Boltzmann law $L = 4 \pi R^2 \sigma T^4$, where $L$ is the luminosity, $R$ is the radius, $\sigma$ is the Stefan–Boltzmann constant, and $T$ is the temperature, assuming that the radiation geometry was a sphere (see Fig. \ref{sfig:SED} for the model fits).
% we derived the radius of the emission region, which we regard as the inner radius of the accretion disk, in units of the Schwarzschild radius for a non-rotating black hole.
Here, we assumed a black hole mass of $\rm 8 \times 10^{8}~M_\odot$ \citep{Scharw2013}. 
As a result, the inner radius of the accretion disk was around 2.4$-$4.1 times the Schwarzschild radius, and the Eddington luminosity ratio, $L/L_{\rm{Edd}}$ was low, around 0.0002$-$0.0005. 
The AGNs and X-ray binaries with such a low Eddington ratio are considered to have a truncated disk with a large inner radius.
Therefore, the shape of the UV peak was not only a blackbody but also had contributions from unresolved line emissions and thermal emission associated with the star-forming activity associated with an infalling galaxy \citep{Yu2015}.
Spectroscopic UV observations are needed to resolve this issue.

In any case, the UV emission from the nucleus of NGC 1275 was not simple, and jet emission should be considered to some extent in modeling the broad-band SED. 
% We further discuss about jet contribution in the next subsection.

\clearpage

% X-ray origin discussion
\subsection{Origin of the X-ray emission}
\label{xray_origin}

Judging from the long-term variability in both the soft and hard X-ray bands, synchronized with the GeV gamma-ray flux increases (Figs. \ref{fig:lightcurve_xrt_uvot} and \ref{fig:bat_comparison}), jet emission is likely to contribute to the X-ray continuum, as reported by \cite{Fukazawa2018}.
In this case, the X-ray emission from the jet is the lowest energy-component of the inverse-Compton scattering radiation in the SSC model.
\cite{Fabian2015} reported a correlated X-ray variability with radio flux from 1980 to 2010 \citep{Trippe2011,ODea1984}, which also supports the contribution of jet emission to the X-ray band.
However, the flux increase from 2013 to 2017 was not as large as that of the GeV gamma-rays, which showed an increase by a factor of three or more.
This is the same for TeV gamma-ray variability \citep{Sinitsyna2020}.
As seen in the spectral variability of the blazars, the low-energy part of the inverse Compton component in the X-ray band does not vary as much as that of the GeV gamma-rays \citep[e.g.][]{Hayashida2012}; thus, this could be a reason for the smaller flux increase in the X-ray band.

\begin{figure}[htbp]
  \begin{center}
    \begin{tabular}{c}

      % PC spectrum
      \begin{minipage}{0.4\hsize}
        \begin{center}
        (a)
        \includegraphics[width=7.cm]{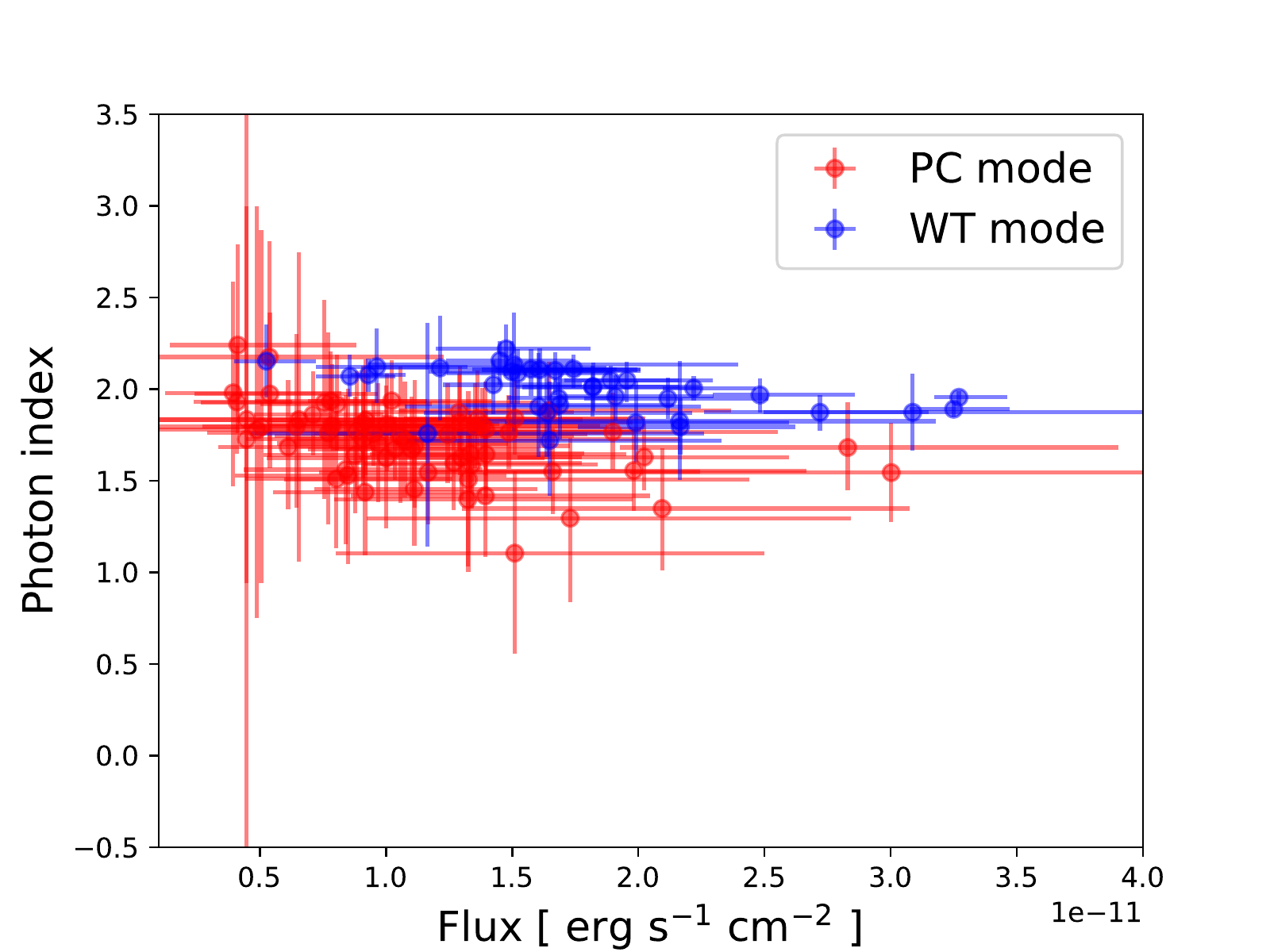}
        \end{center}
      \end{minipage}

      % WT spectrum
      \begin{minipage}{0.4\hsize}
        \begin{center}
        (b)
        \includegraphics[width=7.cm]{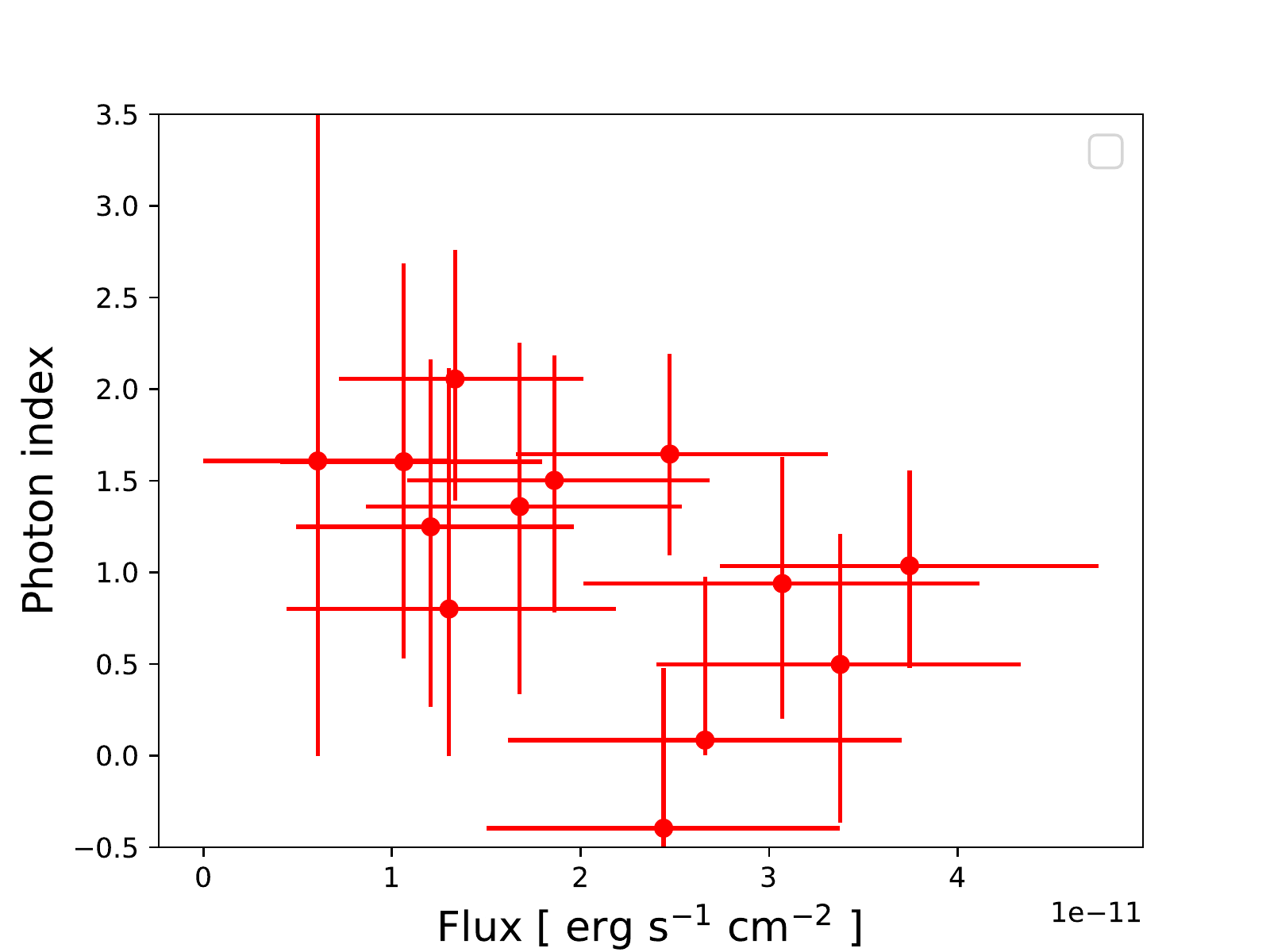}
        \end{center}
      \end{minipage}
      
    \end{tabular}
    \caption{Photon index versus flux figures. (a) {\it Swift}/XRT. Energy range is 5$-$10 keV. The red and blue circles represent the PC- and WT-mode data, respectively. (b) {\it Swift/BAT}. Energy range is 50$-$150 keV.}
    \label{fig:XRT_BAT_flux_and_photonindex}
  \end{center}
\end{figure}

However, like the UV band, short-term soft X-ray variability is not necessarily synchronized with the GeV gamma-rays.
This suggests another origin of the X-ray emission.
Non-beamed emission from a corona is conceivable in addition to jet emission in the soft X-ray band, as inferred from the neutral Fe K$\alpha$ line \citep{Hitomi2018PASJ}. 
This emission is considered to be a fluorescence line coming from the circumnuclear molecular disk irradiated by the bright X-ray continuum from the nucleus. Beamed jet emission cannot irradiate a torus or a disk. 
Disk/corona emission has a similar spectral shape: a power law with a photon index of 1.7$-$2.0, similar to the low-energy tail of the inverse-Compton component of the SSC model \citep{Fukazawa2011}. 
Our results for the soft X-ray photon index are shown in Fig. \ref{fig:XRT_BAT_flux_and_photonindex} (a), indicating that most photon index values were consistent, with a range of 1.7$-$2.0. 
For the BAT data (Fig. \ref{fig:XRT_BAT_flux_and_photonindex} (b)), when the flux became higher, the index became smaller than 1.7.
Although the statistical error was large and we had to consider the systematic error for the BAT data, this might indicate jet emission since such a hard spectrum is not likely to indicate the disk/corona emission in the X-ray band.
In addition, by examining a cutoff in the hard X-ray spectrum, we could examine whether the origin for the X-ray emission was the disk/corona. 
In such a case, the cutoff energy is in the spectrum around 100 keV, whereas the jet emission extends to GeV. 
The accuracy of the BAT (15-150 keV) spectrum (Fig. \ref{fig:bat_comparison}) did not allow us to determine whether there was a cutoff in the hard X-ray spectrum of NGC 1275. 
\citet{Rani2018} also found no clear cutoff in the NuSTAR or BAT data. 
It is thus difficult to separate disk/corona and jet emission from the X-ray spectra alone.
However, there are no clear correlations between the soft X-rays and UV bands (Fig. \ref{fig:correlations} (d)).
This may be a result of the UV emission being contributed by several components in addition to a disk, e.g., line emission, as discussed in \S4.1.
The variability of Fe-K emission can trace the disk/corona emission \citep{Fukazawa2016}.
For NGC 1275, \cite{Hitomi2018PASJ} reported no significant variation of the Fe-K line intensity, regardless of the continuum variability.
This suggests that part of the continuum flux does not contribute to the Fe-K line emission; beamed jet emission is inferred.

The emission site of GeV gamma-ray emission also requires the debate.
No significant changes in the radio band were found after the detection of high gamma-ray activity \citep{Nagai2012}.
VLBI observation revealed two possible gamma-ray emission sites.
One is the radio-brightest C3 component \citep{Nagai2012}, and the other is the C1 component \citep{Nagai2016}.
Additionally, \cite{Hodgson2018} suggested two gamma-ray emission regions based on studies of radio and gamma-ray variability.
Therefore, the origin of variability of the NGC 1275 core is not so simple.

\begin{figure}
    \centering
    \includegraphics[width=15.cm]{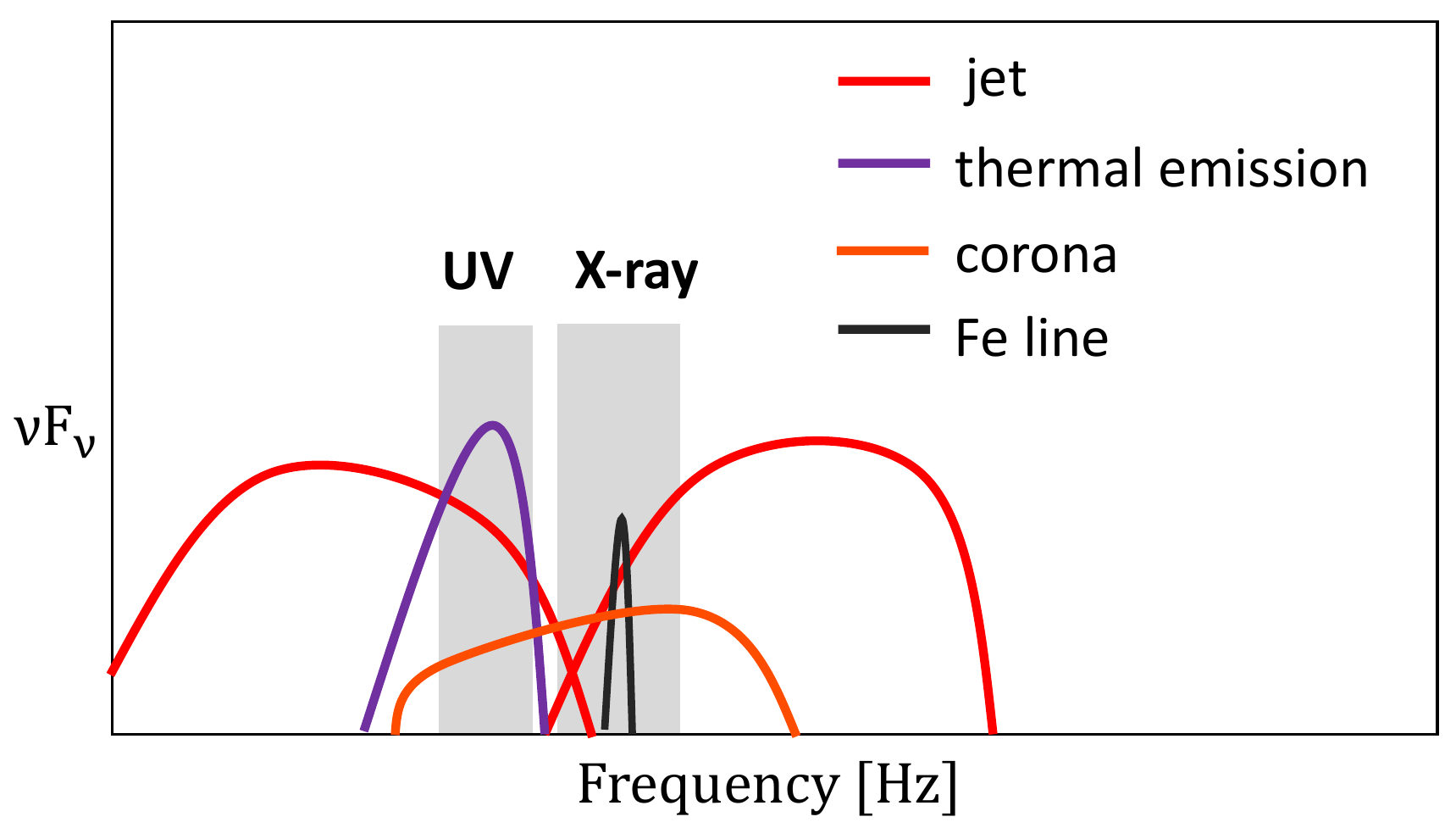}
    \caption{Illustrative contributions of each spectral component to the NGC 1275 SED.}
    \label{fig:predicted_SED}
\end{figure}

As a summary, we illustrate our view of the optical/UV to X-ray emission from the NGC 1275 nucleus in Fig. \ref{fig:predicted_SED}.

% -----

\clearpage
%
%__________________________________________________________________
\section{Conclusions}

Using long-term UV and X-ray data for NGC 1275 obtained over about 10 years with the {\it Swift} satellites UVOT and XRT and the {\it Fermi} satellite LAT, we conducted a variability analysis to investigate the origin of the nuclear emission from NGC 1275 from the UV to the X-ray bands. 
We confirmed that the UV and X-ray fluxes have gradually increased along with the GeV gamma-ray flux.
At times, the UV and X-rays have shown short-term variability associated with a GeV gamma-ray flare.
The UV spectrum can be well represented with a single-temperature blackbody model.
The UV/X-ray and GeV gamma-ray fluxes showed a weak positive correlation, but the UV and X-ray fluxes did not.
The UV color-magnitude diagram showed that the color is almost constant with the changes in magnitude.
As a summary, we show in Fig. \ref{fig:predicted_SED} demonstrates our view of the contributions of each spectral component to the NGC 1275 SED.

% Reference
% the end of the reference

\end{document}